\documentclass[sigconf,screen]{acmart}

\usepackage{caption}
\usepackage[ruled,vlined]{algorithm2e}
\usepackage{multicol}
\usepackage{multirow}
\usepackage{float}
\usepackage{subcaption}
\usepackage[binary-units, group-separator={,}]{siunitx}
\usepackage{pifont}

\newcommand{\StreamData}{\textit{StreamData}\xspace}
\newcommand{\DenseData}{\textit{DenseData}\xspace}
\newcommand{\Type}{\textit{Type}\xspace}
\newcommand{\TypeData}{\textit{TypeData}\xspace}
\newcommand{\StridedBlock}{\textit{StridedBlock}\xspace}

\AtBeginDocument{%
  \providecommand\BibTeX{{%
    \normalfont B\kern-0.5em{\scshape i\kern-0.25em b}\kern-0.8em\TeX}}}

\setcopyright{acmcopyright}
\copyrightyear{2021}
\acmYear{2021}

\acmBooktitle{HPDC '21: ACM Symposium on High-Performance Parallel and Distributed Computing, June 21--25, 2021, Stockholm, Sweeden}

\begin{document}

\renewcommand\_{\textunderscore\allowbreak}

\ifdefined\DRAFT
\newcommand{\cp}[1]{[{\color{green}CP: #1}]}
\newcommand{\wh}[1]{[{\color{red}WH: #1}]}
\newcommand{\ic}[1]{[{\color{blue}IC: #1}]}
\newcommand{\jx}[1]{[{\color{blue}JX: #1}]}
\newcommand{\kw}[1]{[{\color{blue}KW: #1}]}
\else
\newcommand{\cp}[1]{}
\newcommand{\wh}[1]{}
\newcommand{\ic}[1]{}
\newcommand{\jx}[1]{}
\newcommand{\kw}[1]{}
\fi

\title{TEMPI: An Interposed MPI Library with a Canonical Representation of CUDA-aware Datatypes}

\author{Carl Pearson}
\orcid{0001-6481-970X}
\affiliation{%
  \institution{University of Illinois Urbana-Champaign}
  \city{Urbana}
  \state{Illinois}
  \country{USA}
}
\email{pearson@illinois.edu}

\author{Kun Wu}
\affiliation{%
  \institution{University of Illinois Urbana-Champaign}
  \city{Urbana}
  \state{Illinois}
  \country{USA}
}
\email{kunwu2@illinois.edu}

\author{I-Hsin Chung}
\affiliation{%
  \institution{IBM T. J. Watson Research}
  \city{Yorktown Heights}
  \state{New York}
  \country{USA}
}
\email{ihchung@us.ibm.com}

\author{Jinjun Xiong}
\affiliation{%
  \institution{IBM T. J. Watson Research}
  \city{Yorktown Heights}
  \state{New York}
  \country{USA}
}
\email{jinjun@us.ibm.com}

\author{Wen-Mei Hwu}
\affiliation{%
  \institution{Nvidia Research}
  \city{Champaign}
  \state{Illinois}
  \country{USA}
}
\email{whwu@nvidia.com}

\hyphenation{data-types data-type MVA-PICH}

\renewcommand{\shortauthors}{Pearson et al.}

\begin{abstract}
MPI derived datatypes are an abstraction that simplifies handling of non-contiguous data in MPI applications.
These datatypes are recursively constructed at runtime from primitive Named Types defined in the MPI standard.
More recently, the development and deployment of CUDA-aware MPI implementations has encouraged the transition of distributed high-performance MPI codes to use GPUs.
Such implementations allow MPI functions to directly operate on GPU buffers, easing integration of GPU compute into MPI codes.
This work first presents a novel datatype handling strategy for nested strided datatypes, which finds a middle ground between the specialized or generic handling in prior work.
This work also shows that the performance characteristics of non-contiguous data handling can be modeled with empirical system measurements, and used to transparently improve MPI\_Send/Recv latency.
Finally, despite substantial attention to non-contiguous GPU data and CUDA-aware MPI implementations, good performance cannot be taken for granted.
This work demonstrates its contributions through an MPI interposer library, TEMPI.
TEMPI can be used with existing MPI deployments without system or application changes.
Ultimately, the interposed-library model of this work demonstrates MPI\_Pack speedup of up to \num{242000}$\times$ and MPI\_Send speedup of up to \num{59000}$\times$ compared to the MPI implementation deployed on a leadership-class supercomputer.
This yields speedup of more than \num{917}$\times$ in a 3D halo exchange with $3072$ processes.
\end{abstract}


%
%
\maketitle

\section{Introduction}

MPI Derived Datatypes~\cite{mpi31} are an abstraction for describing the layout of non-contiguous data in memory.
They allow MPI functions to operate on such data without intermediate handling by the user application, especially packing the data into a contiguous buffer before transfer.
As GPUs have become a dominant high-performance computing accelerator, MPI implementations such as OpenMPI~\cite{graham2006openmpi}, MVAPICH~\cite{panda2020mvapich}, Spectrum MPI~\cite{spectrum101} and MPICH~\cite{gropp2002mpich} have become ``CUDA-aware''.
In such implementations MPI can directly operate on CUDA device allocations to streamline application development and potentially accelerate inter-process transfers of GPU-resident data.

Previous works have 
created specialized strategies for specific datatypes~\cite{wang2011optimized, shi2014hand},
contributed solutions to handling arbitrary datatypes on GPUs~\cite{jenkins2014processing},
integrated datatype handling with the communication layer~\cite{wei2016gpuaware, hashmi2020falcon},
and tackled the latency of GPU operations~\cite{chu2020dynamic}.
Despite this substantial attention to non-contiguous data and broad GPU deployment in high-performance distributed computing, fast handling of GPU datatypes may not be available in practice.

This work makes three contributions
\begin{itemize}
    \item A novel strategy for converting nested strided MPI datatypes to a compact, canonical representation to enable fast GPU handling (Sec.~\ref{sec:implementation}).
    \item Low-overhead runtime selection of non-contiguous packing strategy for MPI\_Send and MPI\_Recv according to empirical system properties (Sec.~\ref{sec:model}).
    \item A portable implementation based off an MPI interposer library, tested with MVAPICH, Spectrum MPI, and Open MPI on POWER and x86 platforms, and evaluated in-depth against Spectrum MPI on Summit (Sec.~\ref{sec:interpose}).
\end{itemize}

Primarily, this work presents a new approach to handling strided MPI datatypes.
Prior work (Section~\ref{sec:related}) recognizes that MPI datatypes can be generalized to a list of contiguous blocks defined by offsets and sizes.
Further improvements include
specialization for types with certain kinds of regularity~\cite{wang2011optimized,shi2014hand} or 
sophisticated strategies for handling arbitrary datatypes~\cite{jenkins2014processing, chu2019high}.
This work draws a middle ground by observing that compositions of contiguous, vector, hvector, and subarray types are all special cases of an object suitable for compact representation.
A translation phase converts the datatype into an in-memory representation, a canonicalization phase generates a simplified representation, and a parameterized kernel is selected to pack and unpack the data transparently.
This affords wide coverage of structured non-contiguous data without performance fragility of specialized kernels or large metadata sizes for arbitrary datatype handling.

This work also presents a performance model for understanding the impact of packing strategies, and shows it can be used at runtime to improve performance of datatype handling.
Prior work has largely focused on ``one-shot'' methods where non-contiguous GPU data is transferred directly into contiguous host memory through the CUDA ``zero-copy'' mechanism.
This model shows that the one-shot method may not be preferable for non-contiguous data with large strides - depending on the properties of the non-contiguous data and the system, it may be preferable to pack data into GPU memory and use the underlying CUDA-aware mechanism for inter-process transfer.
The model is evaluated in the context of the OLCF Summit system.

Finally, though CUDA-aware MPI implementations have varying degrees of fast non-contiguous data handling, such support cannot be taken for granted for a particular datatype or platform.
The Summit computer at the Oak Ridge Leadership Computing Facility is one such system: it's Spectrum MPI implementation offers functional but extremely slow performance for most MPI datatypes.
The final contribution of this work is to demonstrate that a portable interposed library can transparently deliver large derived-datatype performance improvements without system or application modification.
The Topology Experiments for MPI (TEMPI) library implements this work and has been tested with OpenMPI 4.0.5, MVAPICH 2.3.4, and Spectrum MPI 10.3.1.2.
It transparently converts non-contiguous GPU-resident types to contiguous data via a run-time algorithmic selection before it is passed to the underlying MPI implementation.
TEMPI's interposer design makes it compatible with widely-deployed MPI implementations, but as a consequence it relies on the performance of the contiguous transfer primitives of the underlying MPI implementation.
\ifdefined\BLIND
\else
TEMPI is available at \url{https://github.com/cwpearson/tempi}.
\fi

TEMPI demonstrates a speedup of up to \num{242000}$\times$ for MPI\_Pack and MPI\_Unpack, \num{59000}$\times$ for MPI\_Send, and  \num{20000}$\times$ for a 3D stencil halo exchange on the OLCF Summit system, which does not natively support fast datatype operations on GPU.

This paper is organized in the following way:
Section~\ref{sec:background} introduces MPI derived datatypes, their composition, and the need for compact representation.
Section~\ref{sec:implementation} describes how TEMPI transforms datatypes and selects the kernel.
Section~\ref{sec:interpose} describes the library-interposer method that makes the derived type modifications available without application modification.
Section~\ref{sec:model} describes how datatype-accelerated MPI primitives can be created without system MPI support.
Section~\ref{sec:results} describes the microbenchmark and 3D stencil results.
Section~\ref{sec:future} describes future work for the library.
Section~\ref{sec:related} describes related work.
Finally, Section~\ref{sec:conclusion} concludes.

\section{Composition and Representation of Derived Datatypes}
\label{sec:background}
\begin{figure}[htbp]
     \centering
     \includegraphics[width=\linewidth]{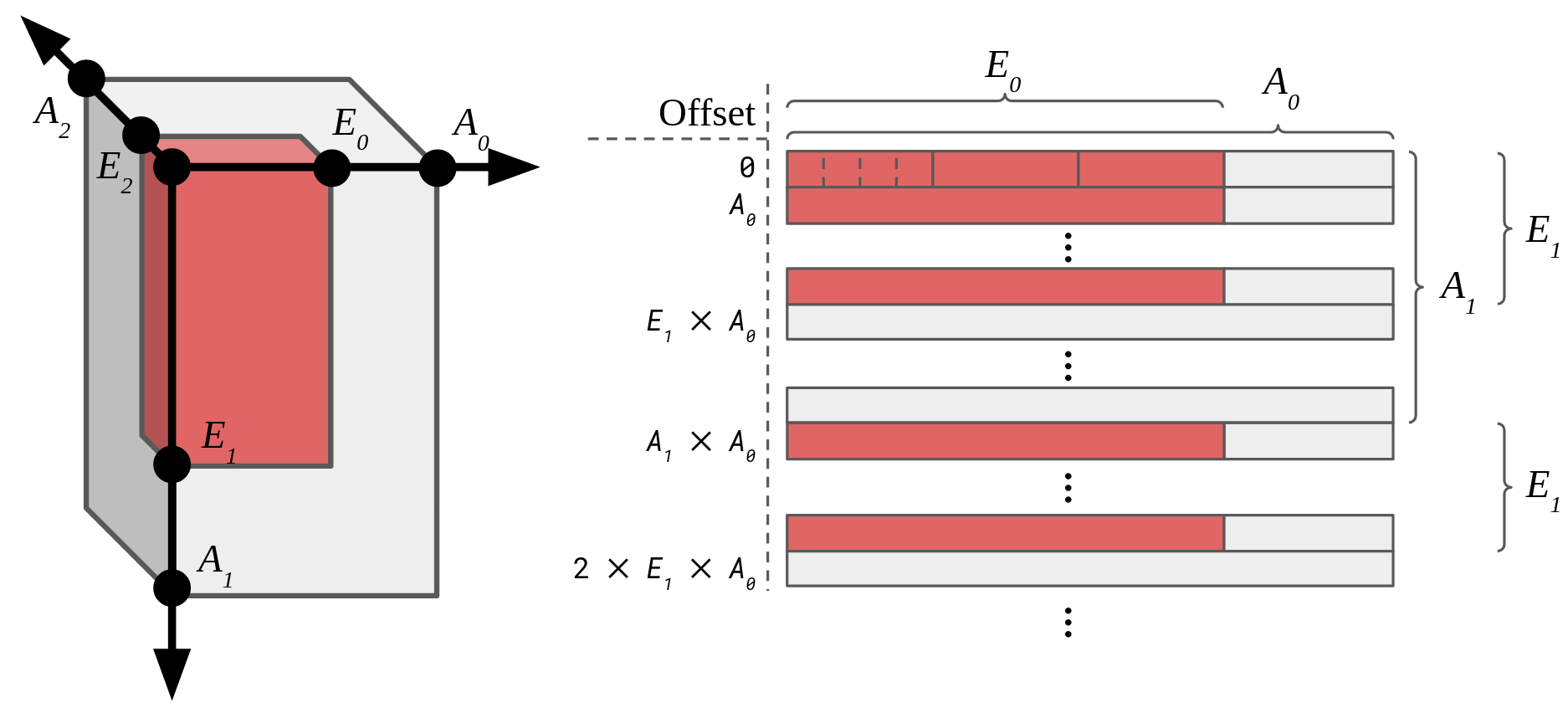}
     \caption{
A 3D object with extent $E_0 \times E_1 \times E_2$ floats in an allocation $A_0 \times A_1 \times A_2$ bytes, and the corresponding linearized memory layout.
         }
     \label{fig:stride}
\end{figure}

Many MPI datatypes can be composed to describe multi-dimensional strided objects.
This work consideres the following ``strided'' datatypes due to their applicability to stencil codes:
\begin{itemize}
\item ``Predefined'' or ``named''\cite[\S3.2.2]{mpi31}: these are the base MPI datatypes (MPI\_BYTE, MPI\_FLOAT, etc.) that correspond to various C or FORTRAN types.
\item ``Contiguous''\cite[\S4.1.2]{mpi31}: these describe ``replication of a datatype in contiguous locations.'' MPI\_Type\_contiguous($n$, \textit{oldtype}, \textit{newtype}): \textit{newtype} is $n$ contiguous repetitions of \textit{oldtype}.
\item ``Vector/Hvector''\cite[\S4.1.2]{mpi31}: these describe ``replication of a datatype into...equally spaced blocks.'' MPI\_Type\_vector($c$, $l$, $s$,  \textit{oldtype}, \textit{newtype}): \textit{newtype} is a vector of $c$ blocks, each block is $l$ contiguous repetitions of \textit{oldtype} and the beginning of each block is separated by $s$ contiguous repetitions of \textit{oldtype}. For hvector, $s$ is given in bytes instead.
\item ``Subarray''\cite[\S4.1.3]{mpi31}: these describe ``n-dimensional subarray of an n-dimensional array.'' MPI\_Type\_create\_subarray($n$, \{\textit{sizes}\}, \{\textit{subsizes}\}, \{\textit{offsets}\} \textit{order}, \textit{oldtype}, \textit{newtype}): \textit{newtype} is an $n$-dimensional subarray of an \textit{oldtype} array with extent   \textit{sizes}. The subarray is of extent \textit{subsizes} at offset \textit{offsets}. \textit{Order} controls C or FORTRAN ordering.
\end{itemize}
These types may be composed in many ways to describe the same non-contiguous bytes.
For example, consider the 3D object in Figure \ref{fig:stride}, which can be visualized as a three-dimensional sub-object of an enclosing three-dimensional object, where the sub-object shares an origin with the enclosing object and each element of the object is a single-precision floating-point number (an MPI\_FLOAT), consuming four bytes.

Each 1D row of the object ($E_0 \times 4$ contiguous bytes) to be described in many ways; a non-exhaustive list follows (meanings of function parameters described in the bulleted list above): 
\begin{itemize}
\item MPI\_Type\_contiguous($E_0$, MPI\_FLOAT, \&row): ``row'' comprises a contiguous replication of $E_0$ single-precision floating-point (4-byte) elements.
\item MPI\_Type\_contiguous($E_0 \times 4$, MPI\_BYTE, \&row): ``row'' is $E_0 \times 4$ 1-byte elements.
\item MPI\_Type\_vector($1$, $E_0$, $1$, MPI\_FLOAT, \&row)
\item MPI\_Type\_vector($E_0$, $4$, $4$, MPI\_BYTE, \&row)
\item MPI\_Type\_create\_hvector($E_0 \times 4$, $1$, $1$, MPI\_BYTE, \&row)
\item MPI\_Type\_create\_subarray($1$, $\{A_0\}$, $\{E_0\}$, $\{0\}$, MPI\_ORDER\_C, MPI\_FLOAT, \&row)
\item MPI\_Type\_create\_subarray($1$, $\{A_0 \times 4\}$, $\{E_0 \times 4\}$, $\{0\}$, MPI\_ORDER\_C, MPI\_BYTE, \&row)
\end{itemize}
These are equivalent for describing a single row, but are not entirely interchangeable since their extents vary.
This distinction is relevant for certain compositions of these types (e.g., below), or when multiple types are manipulated at once.

A 2D plane ($E_1$ rows, offset by $A_0$ bytes between the beginning of each row) can be constructed directly from named types:
\begin{itemize}
\item MPI\_Type\_vector($E_1$, $E_0$, $A_0$, MPI\_FLOAT, \&plane)
\item MPI\_Type\_vector($E_1$, $E_0 \times 4$, $A_0$, MPI\_BYTE, \&plane)
\item MPI\_Type\_create\_subarray($2$, $\{A_0, A_1\}$, $\{E_0, E_1\}$, $\{0, 0\}$, MPI\_ORDER\_C, MPI\_FLOAT, \&plane)
\item MPI\_Type\_create\_subarray($2$, $\{A_0 \times 4, A_1\}$, $\{E_0 \times 4, E_1\}$, $\{0, 0\}$, MPI\_ORDER\_C, MPI\_BYTE, \&plane)
\end{itemize}

or alternatively, as an hvector of rows:

\begin{itemize}
\item MPI\_Type\_create\_hvector($E_1$, $1$, $A_0$, row, \&plane)
\end{itemize}

or for the subarray row types:
\begin{itemize}
\item MPI\_Type\_vector($E_1$, 1, 1, row, \&plane)
\item MPI\_Type\_create\_subarray($1$, ${A_1}$, ${E_1}$, ${0}$, MPI\_ORDER\_C, row, \&plane)
\end{itemize}

Similarly planes comprise a cuboid ($E_2$ planes, offset by $A_0 \times A_1$ bytes between the beginning of each plane).
For example,
\begin{itemize}
\item MPI\_Type\_create\_hvector($E_2$, $1$, $A_0 \times A_1$, plane, \&cuboid)
\item MPI\_Type\_create\_subarray($2$, $\{A_0, A_1, A_2\}$, $\{E_0, E_1, E_2\}$, $\{0, 0, 0\}$, MPI\_ORDER\_C, MPI\_FLOAT, \&cuboid)
\item MPI\_Type\_create\_subarray($2$, $\{A_0 \times 4, A_1, A_2\}$, $\{E_0 \times 4, E_1, E_2\}$, $\{0, 0, 0\}$, MPI\_ORDER\_C, MPI\_BYTE, \&cuboid)
\end{itemize}

In the most general sense, a datatype can be considered as a list of contiguous blocks, where each has an offset and a size.
Indeed, many prior works use such a representation in the general case~\cite{wang2011optimized, shi2014hand, wei2016gpuaware, hashmi2020falcon}, or with additional optimization~\cite{jenkins2014processing, chu2019high}.
The weakness of this approach is that representing datatype may consume as much GPU memory as the datatype itself.

Consider such a representation of $N$ non-contiguous blocks of $M$ MPI\_FLOATs.
To support objects dispersed across large address ranges, the block offset and size would each be $8$ bytes ($64$ bits) each, yielding at least $16 \times N$ bytes to represent $M \times N \times 4$ bytes of data.
If $M$ is relatively small (common for any higher-dimension object) the representation will consume similar memory to the data itself, limiting the space left for the application.

Many works adopt specialized kernels to handle certain common datatypes~\cite{wang2011optimized, shi2014hand, wei2016gpuaware}.
These naturally lend themselves to specific compact representations, e.g. an MPI vector of any size as only a block length, block count, and stride.
Unfortunately, the combinatorial explosion of equivalent representations renders the strategy of specialized kernels infeasible in general.

This work recognizes that structured data in many MPI applications do not require a generic and costly representation.
Compositions of strided datatypes can adequately cover many cases and are amenable to a common compact representation despite the variety of equivalent constructions.
Such a representation uses a negligible amount of GPU memory for each datatype, and minimizes the maintenance burden as a small number of generic packing kernels can cover many datatypes.
Section~\ref{sec:implementation} describes how this is achieved.

\section{MPI Derived Datatype Handling}
\label{sec:implementation}

TEMPI provides a transparent translation from non-contiguous to contiguous data between the application and the MPI implementation.
A packing strategy is created for each datatype the application calls MPI\_Type\_commit on.
When later operations use that datatype, TEMPI first packs the non-contiguous data into a contiguous buffer before passing it on to MPI (and unpacks the data before returning it to the application).
This necessarily places the packing and unpacking operations on the critical path.
This section describes how the packing strategy is selected.

TEMPI uses a three-phase translation/transformation/kernel selection approach to convert distinct-but-equivalent MPI datatypes to a common format.
The MPI\_Type\_commit(\textit{datatype}) function delineates the boundary between when an application constructs a datatype and when that type may be used with the rest of the MPI functions.
The MPI standard advises that ``the system may compile at commit time an internal representation for the datatype \ldots and select the most convenient transfer mechanism.''~\citep[p.~110]{mpi31}.
In line with that advice, TEMPI's phases are implemented within the MPI\_Type\_commit function and cached for later use in MPI functions:
\begin{enumerate}
\item Translation to an internal representation (IR) (Section~\ref{sec:conversion})
\item Transformation to a compact representation (Section~\ref{sec:simplify})
\item Kernel selection and on-device representation (Section~\ref{sec:kernel-selection})
\end{enumerate}

\subsection{Type Translation}
\label{sec:conversion}

The first phase of the datatype handling process is to convert a fully specified MPI derived datatype into a \Type hierarchy, which represents a (possibly non-contiguous) set of bytes from a memory region.
Each \Type level has a field \textit{data} of \TypeData, which represents information about the level.
Each \Type also tracks zero or one child \Type levels.
The \Type hierarchy and its children describe the MPI datatype, where the order of the hierarchy matches the hierarchy of the constructed MPI datatype.

The IR currently includes two kinds of \TypeData: \DenseData for contiguous bytes, and \StreamData for strided patterns of a single child Type.
\DenseData plays the same role as a named type in MPI: it represents a sequence of contiguous bytes and has no children.

\begin{enumerate}
    \item \DenseData
    \begin{enumerate}
        \item integer \textit{offset}, the number of bytes between the lower bound and the first byte of the Type
        \item integer \textit{extent}, the number of contiguous bytes in the Type
    \end{enumerate}
    \item \StreamData, a strided sequence of elements of the child type
    \begin{enumerate}
        \item integer \textit{offset}, as \DenseData
        \item integer \textit{stride}, the number of bytes between elements
        \item integer \textit{count}, the number of elements in the stream
    \end{enumerate}
\end{enumerate}

\begin{figure*}[htbp]
\centering
\includegraphics[width=\textwidth]{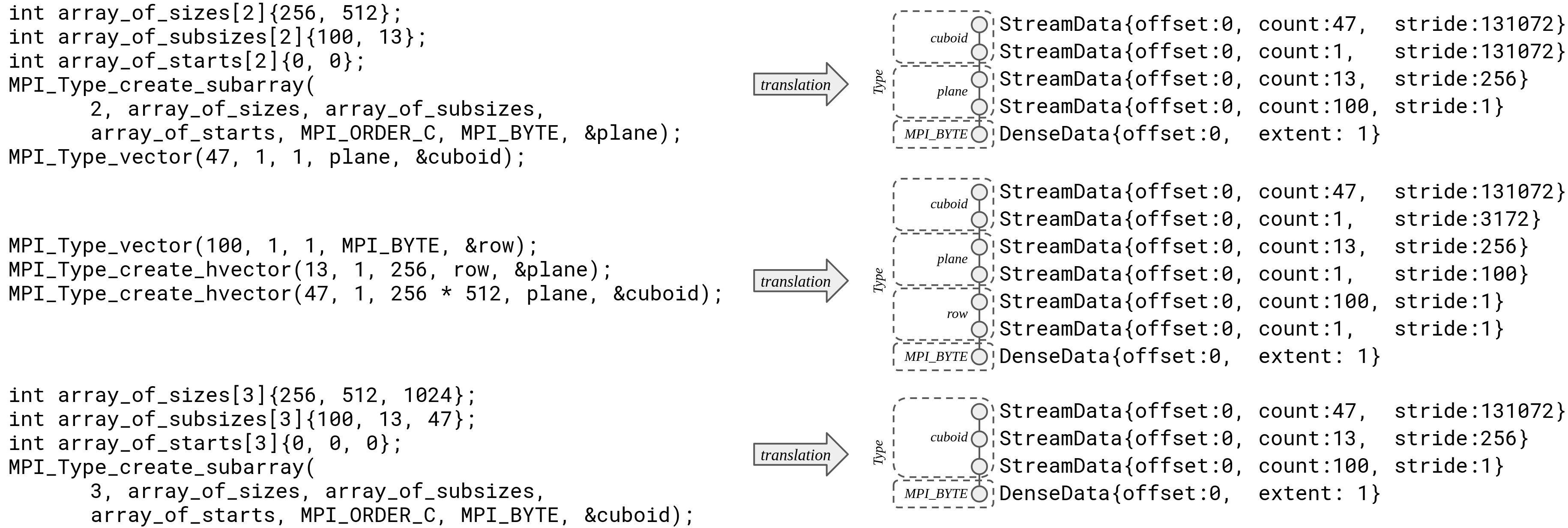}
\caption{
Three different MPI C fragments to generate the 3D object from Fig.~\ref{fig:stride} with $A_0=256$, $A_1=512$, $A_2=1024$, $E_0=100$, $E_1=13$, and $E_2=47$.
The right-hand side shows the corresponding \textit{Type} IR after translation, with parent \TypeData above child \TypeData.
Equivalent objects can be represented differently and require a later transformation pass.
}
\label{fig:translation}
\end{figure*}

Type translation is accomplished by converting each MPI datatype to a corresponding \DenseData or \StreamData node, and then recursively doing the same to its child before attaching them to the converted node.
The recursive base case is when an MPI Named type is reached, which by definition has no children.
Fig.~\ref{fig:translation} shows three different MPI C snippets to create the 3D object described in Fig.~\ref{fig:stride}.

\SetAlFnt{\footnotesize}
\SetKwInOut{Input}{input}
\SetKwInOut{Output}{output}
\SetKwInOut{InOut}{in/out}
\SetKwComment{Comment}{$\triangleright$\ }{}
\SetKw{KwBy}{by}
\SetKw{KwTo}{to}
\SetKw{KwOf}{of}
\SetKw{KwIs}{is}
\SetKw{KwIsNot}{is not}
\SetKw{KwFalse}{FALSE}
\SetKw{KwTrue}{TRUE}
\SetKwProg{Fn}{Function}{:}{}
\SetKwFunction{Fdatatype}{from\_mpi\_datatype}
\SetKwFunction{Fv}{from\_vector}
\SetKwFunction{Fhv}{from\_hvector}
\SetKwFunction{Fc}{from\_contiguous}
\SetKwFunction{Fna}{from\_named}
\SetKwFunction{Fs}{from\_subarray}
\SetKwFunction{Fsimplify}{simplify}
\SetKwFunction{Fstreamelision}{stream\_elision}
\SetKwFunction{Fstreamflatten}{stream\_flatten}
\SetKwFunction{Fdensefolding}{dense\_folding}
\SetKwFunction{Fsort}{sort}
\SetKwFunction{Fstridedblock}{strided\_block}

An MPI named type (MPI\_INT, etc.) is translated into a \DenseData with the \textit{extent} field equal to the extent of the named type, and offset $0$.
A named type is not a derived type, so it has no children.

An MPI contiguous type (MPI\_Type\_contiguous) is a special case of \StreamData where the stride matches the size of the element.
It is not \DenseData as \textit{oldtype} may not be dense.
\textit{Offset} is $0$, \textit{stride} equal to the extent of the \textit{oldtype} argument, and \textit{count} equal to the \textit{count} argument.

An MPI vector (MPI\_Type\_vector) or hvector (MPI\_Type\_create\_hvector) are translated into two nested \StreamData, a ``parent'' and ``child''.
The parent represents the repeated blocks, and the child the repeated elements within each block.
Both offsets are $0$.
The child count is the vector blocklength, and the child stride is the extent of \textit{oldtype}.
The parent count is the vector count, and the parent stride is the child \textit{stride} times the vector stride.
For hvector the parent \textit{stride} is given directly in the hvector \textit{stride} argument and does not need to be computed.

An MPI subarray (MPI\_Type\_create\_subarray) is a set of nested \StreamData equivalent to the dimension of the subarray.
MPI subarray arguments are provided inner-to-outer, which corresponds to a descendant-ancestor relationship in the \textit{Type} tree.
The count of dimension $i$ is provided by the corresponding subarray \textit{subsize}.
The stride of dimension $i$ is the product of the MPI extent of the subarray \textit{oldtype} and the $i-1$ preceding subarray sizes.
The offset of each dimension is given in terms of elements and is converted to bytes for the \TypeData.

\subsection{Type Canonicalization}
\label{sec:simplify}

The construction of the \Type hierarchy described in Section \ref{sec:conversion} yields a hierarchy of \textit{StreamData} with a base of \DenseData.
Since each level of the datatype has a direct correspondence in the \Type hierarchy, semantically equivalent datatypes may have different {\Type}s.
In order to provide fast handling of equivalent types, these various representations are canonicalized.

Four transformations are used to canonicalize the Type tree.
``Dense folding'' collapses \DenseData into a parent \StreamData.
``Stream elision'' removes a \StreamData representing a stream of one element.
``Stream flattening'' combines two \StreamData that could be represented as one.
``Sorting'' ensures the \StreamData have a unique order.
The optimizations are applied repeatedly in turn, only terminating when neither optimization would modify the \Type hierarchy.
Algorithm~\ref{alg:simplify} summarizes the overall simplification process.

\begin{algorithm}[ht]
\SetAlgoLined
\DontPrintSemicolon
\Fn{\Fsimplify{ty}}{
 simplified $\gets$ ty \\
 changed $\gets$ \KwTrue \\
\While{changed}{
  changed $\gets$ \KwFalse $\lor$ \Fdensefolding(simplified)%
  \Comment*[r]{in-place}
  changed $\gets$ changed $\lor$ \Fstreamelision(simplified)%
  \Comment*[r]{in-place}
  changed $\gets$ changed $\lor$ \Fstreamflatten(simplified)%
  \Comment*[r]{in-place}
  changed $\gets$ changed $\lor$ \Fsort(simplified)%
  \Comment*[r]{in-place}
}
\Return ty\;
}
 \caption{simplification}
 \label{alg:simplify}
\end{algorithm}

\subsubsection{Dense Folding}

Dense folding is driven by the observation that stride of a \StreamData may match the extent of a child \DenseData.
Such a configuration represents a stream of repeated contiguous dense elements.
In that case, the \DenseData extent can be ``folded'' up into the \StreamData, and the pair can be represented as a single \DenseData node.
This scenario may arise when an MPI vector, subarray, or contiguous type is used to describe a contiguous region larger than any MPI named type.

\begin{figure}[ht]
\centering
\includegraphics[width=\linewidth]{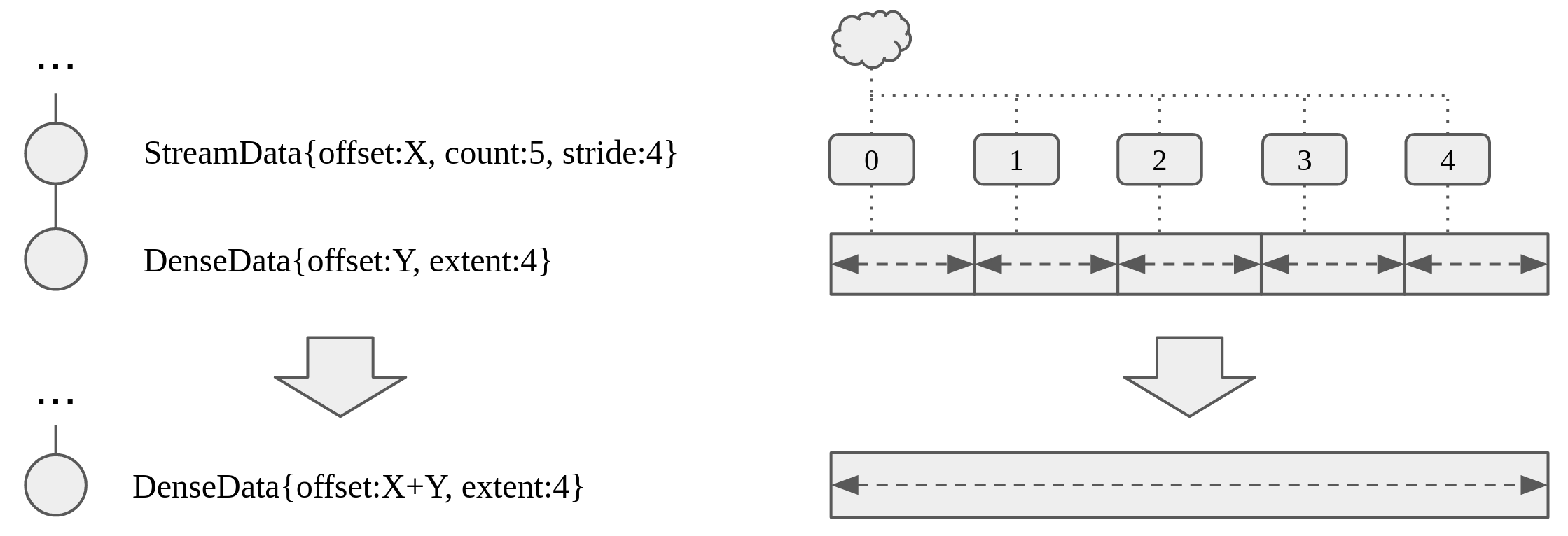}
\caption{
Example of Dense Folding.
When the extent of a DenseData matches the stride of a parent StreamData, the parent/child combination can be replaced with a single larger DenseData.
}
\label{fig:dense-folding}
\end{figure}

Algorithm~\ref{alg:dense-fold} shows how the transformation is applied to a \Type, and Figure \ref{fig:dense-folding} shows the transformation graphically.
The transformation is applied to each Type node of the Type tree in a depth-first order.
At each node, the transformation only applies if the node (\textit{ty}) is a \StreamData kind and the node's child (\textit{child}) is a \DenseData.
If the parent's \textit{stride} matches the child's \textit{extent}, the parent is replaced with a larger \DenseData node that represents the entire contiguous stream.
The child's offset is increased to include any offset the parent had.

\begin{algorithm}[htbp]
\SetAlgoLined
\DontPrintSemicolon
\Fn{\Fdensefolding{ty}}{
 changed $\gets$ \KwFalse\;
 \For{child \KwOf ty} {
   changed $\lor$ \Fdensefolding{child}%
   \Comment*[r]{fold from bottom up}
 }

 \If{ty.data \KwIsNot StreamData} {
 \Return changed\;
 }
 
 Type child = ty.children[0]\;
 \If{child.data \KwIsNot DenseData} {
 \Return changed\;
 }
 
 StreamData cData $\gets$ child.data\;
 StreamData pData $\gets$ ty.data\;
 
 \If{cData.extent == pData.stride}{
   changed $\gets$ \KwTrue\;
   cData.off $\gets$ cData.off $+$ pData.off\;
   cData.extent $\gets$ pData.count $\times$ pData.stride\;
   ty $\gets$ child\Comment*[r]{replace ty with child}
 }
 
 \Return changed
 }
 \caption{dense\_folding from Alg.~\ref{alg:simplify}}
 \label{alg:dense-fold}
\end{algorithm}

\subsubsection{Stream Elision}

\textit{Stream elision} canonicalizes a case where a stream has only a single element.
Consider \textit{ty}, a \StreamData with a child \StreamData whose count \textit{count} is one.
In such a case, \textit{child} is a single element and can be elided.
This construction arises in the case of an MPI vector with \textit{blocklength} 1 dimension with \textit{subsize} 1.

\begin{figure}[htbp]
\centering
\includegraphics[width=\linewidth]{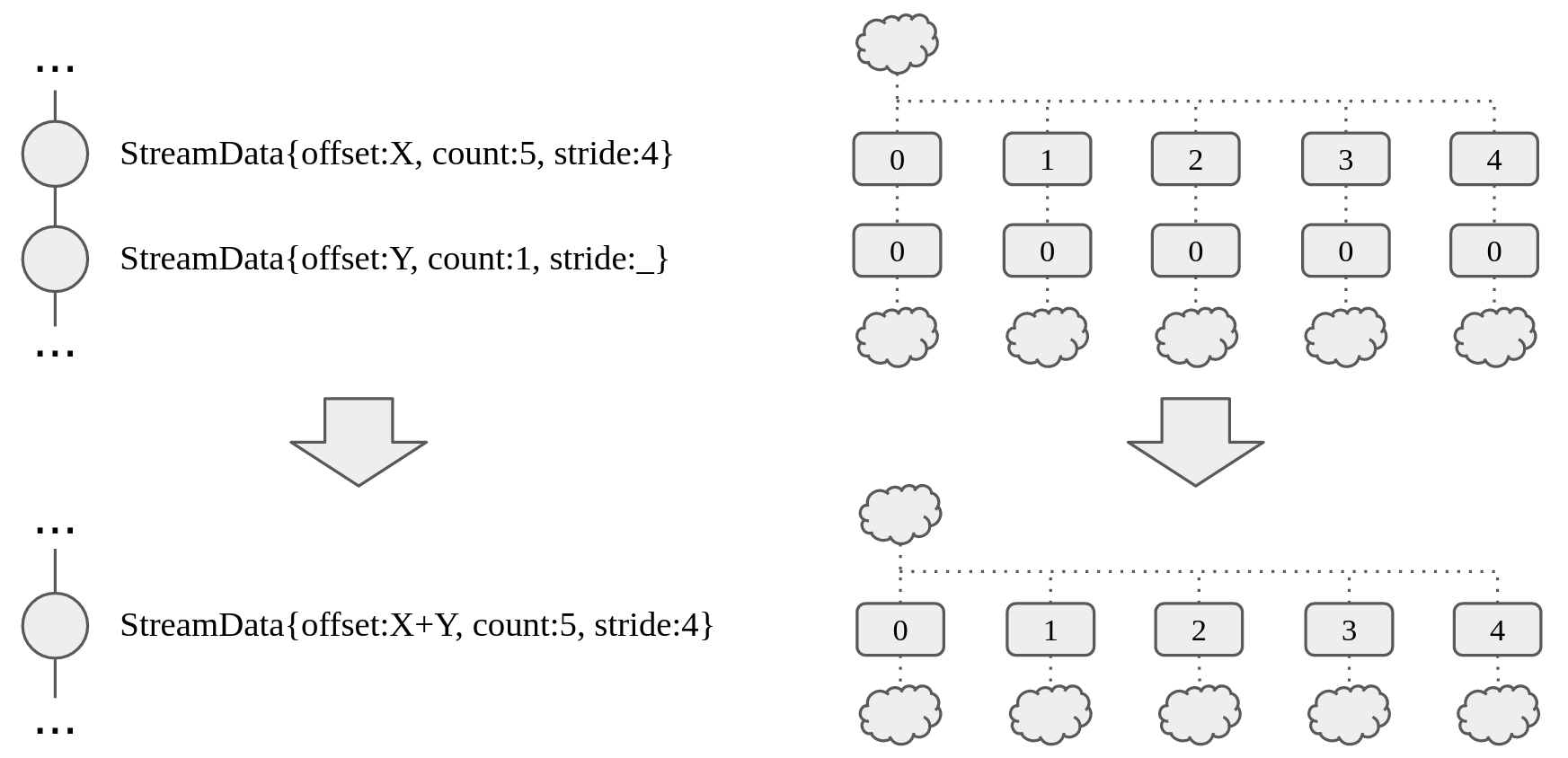}
\caption{
Example of stream elision.
When a child \StreamData has only a single element, it can be removed from the \textit{Type} tree.
}
\label{fig:stream-elision}
\end{figure}

Algorithm~\ref{alg:elide} shows how the transformation is applied to a \textit{Type}, and Figure \ref{fig:stream-elision} shows an example.
Like with dense folding, stream elision is applied separately to each \Type node in a depth-first order.
After that, if both the type \textit{ty} and its child \textit{child} are \StreamData, then if the child has count of $1$, the child is replaced with its own children.

\begin{algorithm}[ht]
\SetAlgoLined
\DontPrintSemicolon
\Fn{\Fstreamelision{ty}}{
 
 changed $\gets$ \KwFalse\;
 \For{child \KwOf ty} {
   changed $\gets$ changed $\lor$ \Fstreamelision{child}%
   \Comment*[r]{bottom up}
 }

 \If{ty.data \KwIsNot StreamData} {
 \Return changed\;
 }
 
 Type child = ty.child\;
 \If{child.data \KwIsNot StreamData} {
 \Return changed\;
 }
 
 StreamData cData $\gets$ child.data\;
 
 \If{1 == cData.count}{
   changed $\gets$ \KwTrue\;
   ty.child $\gets$ child.children\Comment*[r]{delete child}
 }
 
 \Return changed
 }
 \caption{stream\_elision from Alg.~\ref{alg:simplify}}
 \label{alg:elide}
\end{algorithm}

\subsubsection{Stream Flattening}
\textit{Stream flattening} canonicalizes the case where a pair of nested streams could be represented as a single stream.
Consider a child \StreamData with a count $A$ and a stride $B$.
If the parent \StreamData has a stride that is a product of A and B, that means multiple children are separated by the child's stride.
In such a case, the parent and child can be flattened into a single \StreamData with a larger count.

\begin{figure}[ht]
\centering
\includegraphics[width=\linewidth]{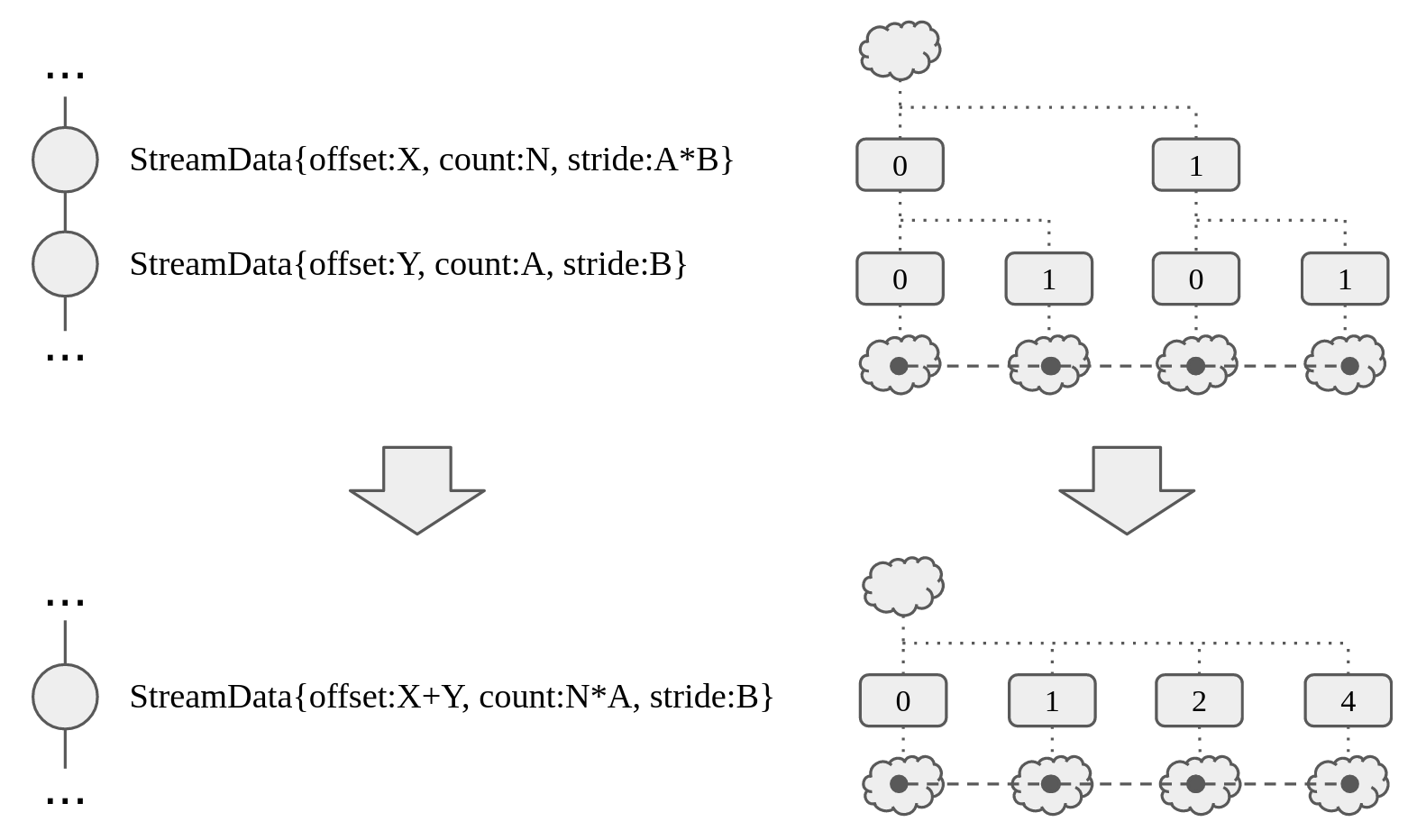}
\caption{
Example of stream flattening.
When the parent stride allows repeated children to maintain a fixed stride between their elements, the parent and child can be flattened into a single stream.
}
\label{fig:stream-flatten}
\end{figure}

Algorithm~\ref{alg:flatten} shows how the transformation is applied to a \textit{Type}, and Figure \ref{fig:stream-flatten} shows an example.
This operation has some overlap with stream elision.
Stream elision handles the specific case when the child's count is 1, which lifts the restriction on the child and parent stride relationship in stream flattening.

\subsubsection{Sorting}
\textit{Sorting} canonicalizes the ordering of a pair of nested streams is arbitrary.
For example, consider a 2D non-contiguous object.
That object could be constructed as columns of rows of blocks, or rows of columns of blocks.
To canonicalize this case, the \StreamData hierarchy is sorted by stride, with the largest strides first in the hierarchy and the smaller strides last.

\begin{algorithm}[ht]
\SetAlgoLined
\DontPrintSemicolon
\Fn{\Fstreamflatten{ty}}{
 
 changed $\gets$ \KwFalse\;
 \For{child \KwOf ty} {
   changed $\gets$ changed $\lor$ \Fstreamflatten{child}%
   \Comment*[r]{bottom up}
 }

 \If{ty.data \KwIsNot StreamData} {
 \Return changed\;
 }
 
 Type child = ty.child\;
 \If{child.data \KwIsNot StreamData} {
 \Return changed\;
 }
 
 StreamData pData $\gets$ ty.data\;
 StreamData cData $\gets$ child.data\;
 
 \If{pData.stride == cData.count $\times$ cData.stride}{
   changed $\gets$ \KwTrue\;
   pData.count $\gets$ pData.count $\times$ cData.count\;
   pData.stride $\gets$ cData.stride\;
   pData.off $\gets$ pData.off + cData.off\;
   ty.child $\gets$ child.children\Comment*[r]{delete child}
 }
 
 \Return changed
 }
 \caption{stream\_flatten from Alg. \ref{alg:simplify}}
 \label{alg:flatten}
\end{algorithm}

\subsection{Kernel Selection}
\label{sec:kernel-selection}

Once the type is canonicalized, it is converted into a \StridedBlock structure.
The \StridedBlock structure is semantically similar to an MPI subarray and is used only to select the kernel implementation.

\begin{itemize}
    \item \StridedBlock
    \begin{itemize}
        \item integer \textit{start}: byte offset between the lower bound and the first element
        \item integer list \textit{counts}: number of elements in the dimension
        \item integer list \textit{strides}: bytes between the start of each element in the dimension
    \end{itemize}
\end{itemize}

The \textit{start} field describes the offset of the first byte in the object from the beginning of the allocation.
The $i$th entry of \textit{counts} and \textit{strides} describes the number of repetitions of the previous dimension and the number of bytes separating each repetition, respectively.

Algorithm~\ref{alg:strided-block} describes the conversion from \Type to \StridedBlock.
This is only possible if the bottom is a DenseData and every other object is a StreamData.
The process in Section \ref{sec:simplify} ensures that structure, if it is possible.
The DenseData describes the first dimension, which will have stride $1$ and count equal to the extent of the DenseData.
Each higher dimension directly corresponds to the StreamData.
The offset of each dimension is accumulated into the single offset of the StridedBlock.

\begin{algorithm}[htbp]
\SetAlgoLined
\DontPrintSemicolon
\Fn{\Fstridedblock{ty}}{
 
 datas $\gets$ []\;
  
 cur $\gets$ ty\Comment*[r]{Add all TypeData to an array}
 \While{true}{
   datas.append(cur)\;
   \eIf{cur.child == \{\}} {
     break%
     \Comment*[r]{no children left}
   }{
     cur $\gets$ cur.child\;
   }
 }
 StridedBlock sb%
 \Comment*[r]{to be returned}
 \For{i = 0 \KwTo datas.size()} {
   \eIf{i == 0} {
     \eIf{data \KwIs DenseData} {
       sb.off $\gets$ data.off\;
       sb.counts.append(data.extent)\;
       sb.strides.append($1$)\Comment*[r]{DenseData stride is 1}
     }{
       \Return NULL\Comment*[r]{Not strided}
     }
   }{
    \eIf{data \KwIs StreamData} {
       sb.off $\gets$ sb.off $+$ data.off\;
       sb.counts.append(data.count)\;
       sb.strides.append(data.stride)\;
     }{
       \Return NULL\Comment*[r]{Not strided}
     }
   }
 }
 \Return sb\;
 }

 \caption{conversion of \textit{Type} to \StridedBlock}
 \label{alg:strided-block}
\end{algorithm}

Once the \textit{Type} is converted into a \textit{StridedBlock}, the next task is to choose a method for fast packing and unpacking on the GPU.
If the StridedBlock is 1D (contiguous), we issue a single cudaMemcpyAsync to move the data into the destination buffer, followed by a cudaStreamSynchronize.
This is similar to the implementation in MVAPICH, OpenMPI, and Spectrum MPI.
If the StridedBlock is 2D we select a kernel that maps the X dimension of the thread index into the count[0] and the Y dimension to count[1].
If the StridedBlock is 3D, we map the X dimension to the count[0], Y dimension to the count[1], and Z dimension to the count[2].
Higher dimensional objects can follow the same general pattern, with additional outer loops for each dimension.

Each kernel dimension is filled from X to Z by the smallest power of two that encompasses the corresponding extent, ultimately limited by a block limit of 1024 threads.
The grid is then sized to cover the entire input object once the block size is determined.

Each kernel is specialized to a word size $W$, which is the largest GPU-native type that is both aligned to the object and is a factor of count[0].
The X dimension collaboratively loads count[0] contiguous bytes that make up each block using elements of size $W$.

Many MPI functions that operate on datatypes accept a \textit{count}, \textit{incount}, or \textit{outcount} parameter, describing how many objects are to be operated on in the buffer.
Unlike other properties of the type, this value is not known until the MPI function is called and therefore is not included in the type optimization.
The kernels handle this value dynamically either by increasing the grid Z dimension (for 2D), or by applying the entire kernel grid to each object in turn (3D).

By the end of this whole process, each MPI datatype has a corresponding kernel implementation with a specific $W$ instantiation.
No metadata is consumed in GPU memory - all object parameters are either encoded into the kernel binary ($W$) or passed as a scalar kernel argument.

\section{Modeling MPI Primitives with Datatype Acceleration}
\label{sec:model}
The interposer design (Section \ref{sec:interpose}) requires that interprocess communication is handled by the underlying system MPI.
Therefore, integration of datatype handling with underlying communication is restricted to packing and unpacking non-contiguous data into contiguous buffers, upon which system MPI primitives are invoked.

In the ``device'' packing method ($T_{device}$, Eq.~\ref{eq:device}), the strided object is packed from the original GPU buffer into an intermediate GPU buffer ($T_{gpu-pack})$, then transferred to an intermediate buffer on the destination GPU with CUDA-aware MPI\_Send/MPI\_Recv ($T_{gpu-gpu}$), then unpacked into the strided destination object ($T_{gpu-unpack}$).

\begin{equation}
T_{device} = T_{gpu-pack} + T_{gpu-gpu} + T_{gpu-unpack}
\label{eq:device}
\end{equation}

In the ``one-shot'' packing method ($T_{oneshot}$, Eq.~\ref{eq:oneshot}), the strided object is packed from the original GPU buffer into intermediate mapped CPU buffer ($T_{host-pack})$, transferred to an intermediate mapped buffer at the destination ($T_{cpu-cpu}$), then unpacked directly into GPU memory ($T_{host-unpack}$).

\begin{equation}
T_{oneshot} =  T_{host-pack} + T_{cpu-cpu} + T_{host-unpack}
\label{eq:oneshot}
\end{equation}

Finally, in the ``staged'' method ($T_{staged}$, Eq.~\ref{eq:staged}) matches the device method, except the intermediate GPU buffer is transferred to a pinned buffer on the host ($T_{h2d}$), where it is transferred to the destination rank's CPU before being copied to the destination GPU ($T_{h2d}$).
This method would only be faster than the device method if $T_{cpu-cpu} + T_{h2d} + T_{d2h} < T_{gpu-gpu}$. 

\begin{equation}
T_{staged} = T_{gpu-pack} + T_{d2h} + T_{cpu-cpu} + T_{h2d} + T_{gpu-unpack}
\label{eq:staged}
\end{equation}

Wang et al.~\cite{wang2011optimized} introduces the one-shot and staged methods (using cudaMemcpy2DAsync instead of GPU kernels).
They find that the staged method is preferable to one-shot.
In contrast, the other works described in Section~\ref{sec:related} prefer the one-shot method with various GPU kernels.

Modeling the performance of each is a challenge in its own right.
Inter-node message latency ($T_{cpu-cpu}$) is commonly modeled as a latency term plus a bandwidth term~\cite{bar1992designing}, possibly refined into short, eager, and rendezvous regimes.
Inter-node GPU message latency ($T_{gpu-gpu}$) further complicates the model with GPU-CPU bandwidth, GPU control latency, direct communication between GPU and NIC (Nvidia's ``GPUDirect''), and pipelining of large messages~\cite{bienz2020modeling}.
When datatypes are involved, there is additional complexity regarding efficiency of non-contiguous memory accesses served through device memory ($T_{gpu-pack}$) or over the CPU-GPU interconnect ($T_{cpu-pack}$).
The interposer design places TEMPI at the mercy of the performance characteristics of the underlying system, so this work sidesteps these concerns by measuring the relevant performance directly and using them at runtime to choose the packing method (Section \ref{sec:send}).

\section{Library Architecture}
\label{sec:interpose}

\begin{figure}[htbp]
         \centering
         \includegraphics[width=\linewidth]{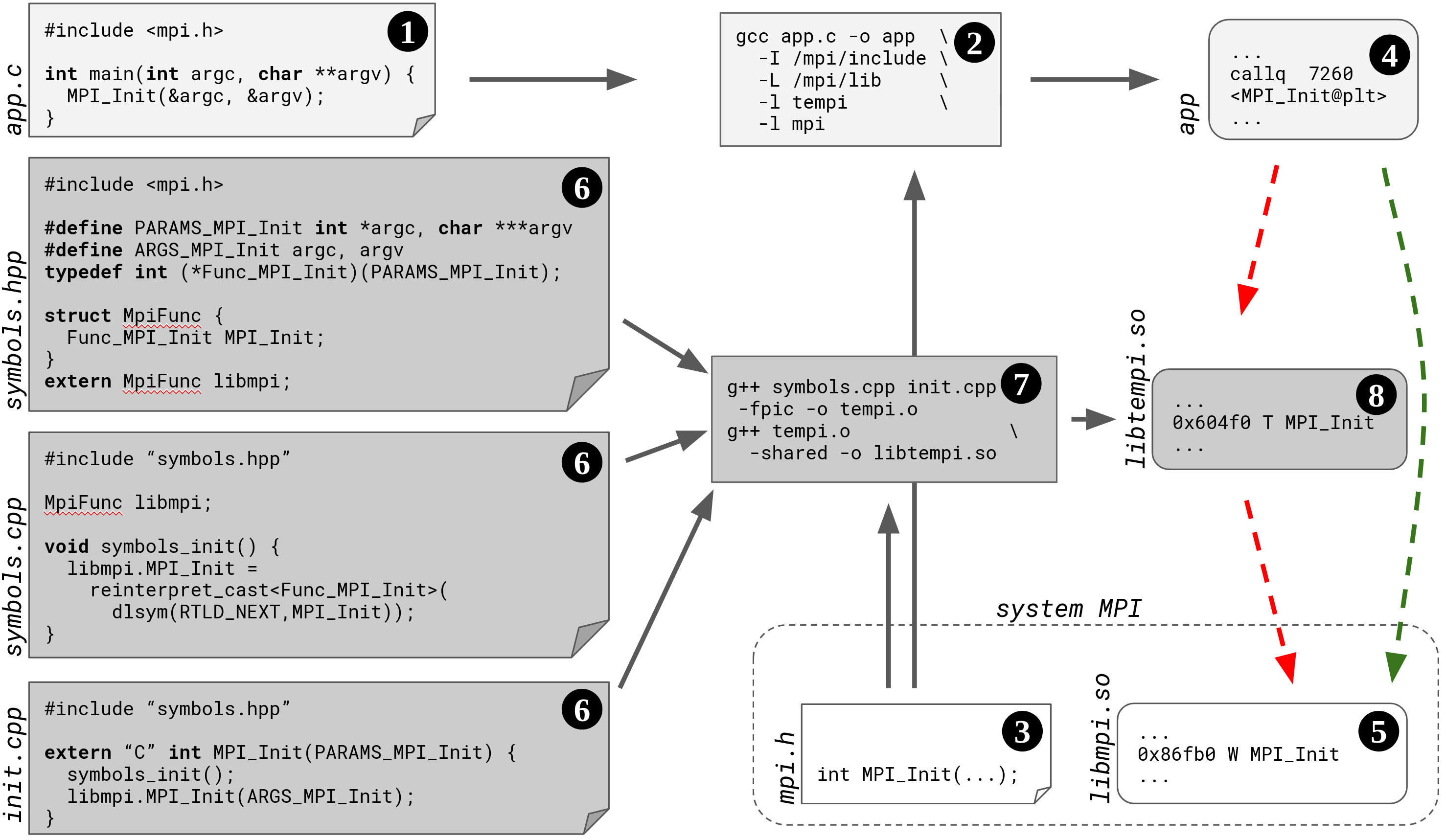}
         \caption{
The application source file (\ding{202}) includes the MPI header file provided by the system (\ding{204}).
It is compiled (\ding{203}) and linked with the system MPI implementation.
When the binary (\ding{205}) is executed, symbols are resolved and the MPI code from the system MPI library is executed.
The TEMPI source files (\ding{207}) are compiled (\ding{208}) into a dynamic library (\ding{209}) using the system MPI header.
The application is compiled as normal except for TEMPI being inserted into the link order (\ding{203}), or an unmodified application can be used with the LD\_PRELOAD mechanism.
When the application is executed, any symbols defined by the TEMPI library will be resolved there (\ding{209}), allowing the TEMPI code to be executed.
Any others will be resolved in the system implementation.
         }
         \label{fig:tempi}
     \end{figure}

The Topology Experiments for MPI library is designed to make MPI modifications available to research and production code without relying on updates to the system MPI implementation.
For reference, Fig.~\ref{fig:tempi} shows a compiled MPI application (\ding{202}-\ding{206}) and the TEMPI interposer (\ding{207}-\ding{209}).
The application source (\ding{202}) includes the system MPI headers (\ding{205}) and is compiled (\ding{203}) to produce a binary (\ding{204}).
At run time, the operating system will resolve the symbols in the application binary according to the order of linked libraries, and MPI\_Init is found in the system MPI implementation (\ding{206}).

TEMPI provides new MPI functionality for unmodified applications by exporting a partial implementation for the MPI interface.
For example, init.cpp (\ding{207}) implements the MPI\_Init function.
The TEMPI source includes the system MPI header, and must be compiled (\ding{208}) with the same MPI as the target application so that the ABI matches.
If the original application can be recompiled, the TEMPI library (\ding{209}) may be inserted into the link order before the system MPI library (\ding{203}).
If not, the TEMPI library can be injected using LD\_PRELOAD or similar mechanism (not shown).

Either way, the operating system will search for the MPI\_Init symbol in the TEMPI library.
As it is found there, that function will be called instead of the system MPI.
Internally, TEMPI may ultimately call some system MPI function after introducing its own functionality.
This is achieved through the dlsym function.
Any parts of the MPI interface that TEMPI does not cover will fall back to the system MPI library automatically.

A transparent transformation from non-contiguous application data to contiguous data provided to MPI introduces some engineering challenges not discussed in detail in this work.
Generally, the performance modeling (described later) as well as CUDA APIs to provide streams, pinned host buffers, devices buffers and events introduce too much latency.
TEMPI uses a caching layer to accelerate repeated requests for the same resources, which is common in iterative applications.
This allows otherwise expensive resources and decisions to be provided tens or hundreds of nanoseconds amortized time, instead of microseconds to milliseconds.

\section{Results}
\label{sec:results}

The experiments are carried out on the OLCF Summit platform, using Spectrum MPI 10.3.1.2.
Summit nodes have two IBM POWER 9 CPUS and six Nvidia V100 GPUs connected in two quadruplets by 100 GB/s bidirectional Nvlink 2 interconnects.
Experiments were performed with NVCC 11.0.221, GCC 9.3.0, and GPU driver 418.116.00.

\subsection{MPI\_Type\_commit}

\begin{figure}[ht]
\centering
\includegraphics[width=\linewidth]{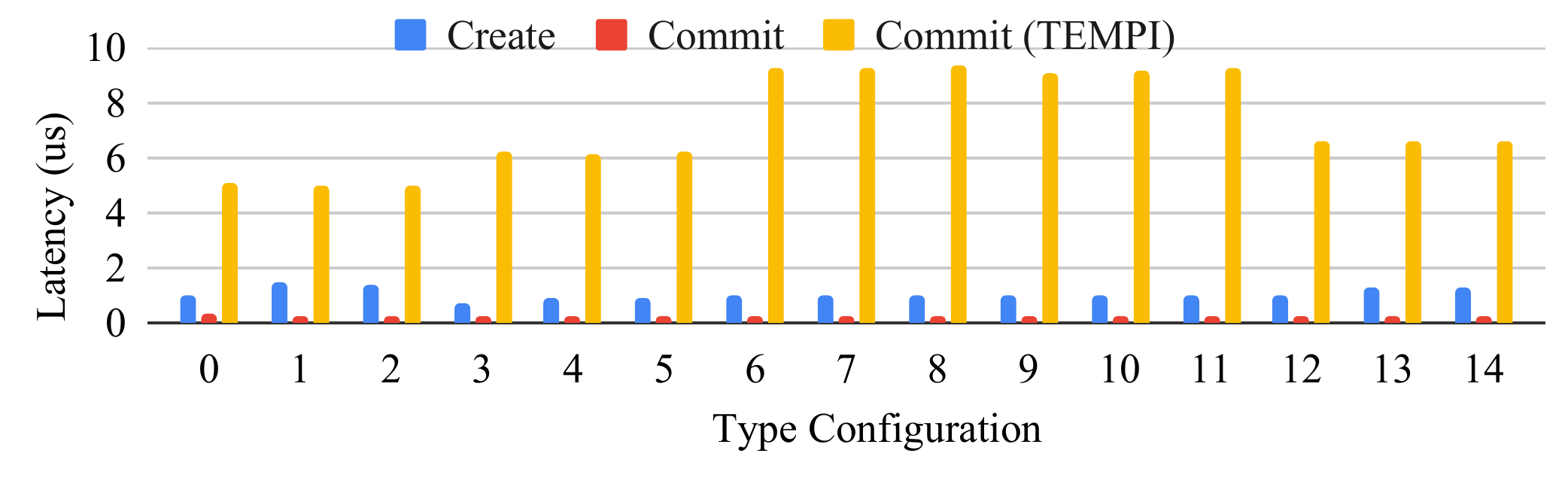}
\caption{
Time for MPI derived datatype creation and commit time for a variety of 3D objects described with subarray (0-2), hvector of vector (3-5), hvector of hvector of vector (6-11), and subarray of vector (12-14).
The ``create'' component uses MPI\_Type* and MPI\_Type\_create* family of MPI APIs to describe the type.
The ``commit'' component is how much time is consumed in MPI\_Type\_commit.
The trimean of $30000$ executions of each phase is reported.
Create time is unchanged (TEMPI does nothing) and is reported for comparison.
TEMPI's performance varies due to different required canonicalization operations.
TEMPI slows commit time substantially, but it still has a negligible impact on application run time.
}
\label{fig:type-commit}
\end{figure}

The type transformation and kernel selection process is executed when the application calls MPI\_Type\_commit.
Fig.~\ref{fig:type-commit} shows the run-time impact of creating MPI derived types, broken down into two phases.
Creation refers to using the MPI\_Type* and MPI\_Type\_create* functions to assemble the type description.
Commit refers to calling MPI\_Type\_commit on that description.
Different type configurations have different commit times as a different sequence of transformations is required to arrive at the canonical form.
Overall, the transformation and kernel selection process slows down the create+commit process by $3.8\times$ to $8.3\times$ on Summit.
This slowdown is a one-time cost during program startup and is small in magnitude.

\subsection{MPI\_Pack and MPI\_Unpack}
\label{sec:pack}

Once a type has been committed, it can be used in a communication routine.
The simplest examination of such a routine is MPI\_Pack, where a buffer is ``sent'' into another buffer in the same process.
When GPU buffers are passed to the \textit{inbuf} and \textit{outbuf} parameters of MPI\_Pack/Unpack, TEMPI uses the selected GPU kernel to complete the packing.

\begin{figure}[ht]
\centering
\includegraphics[width=\linewidth]{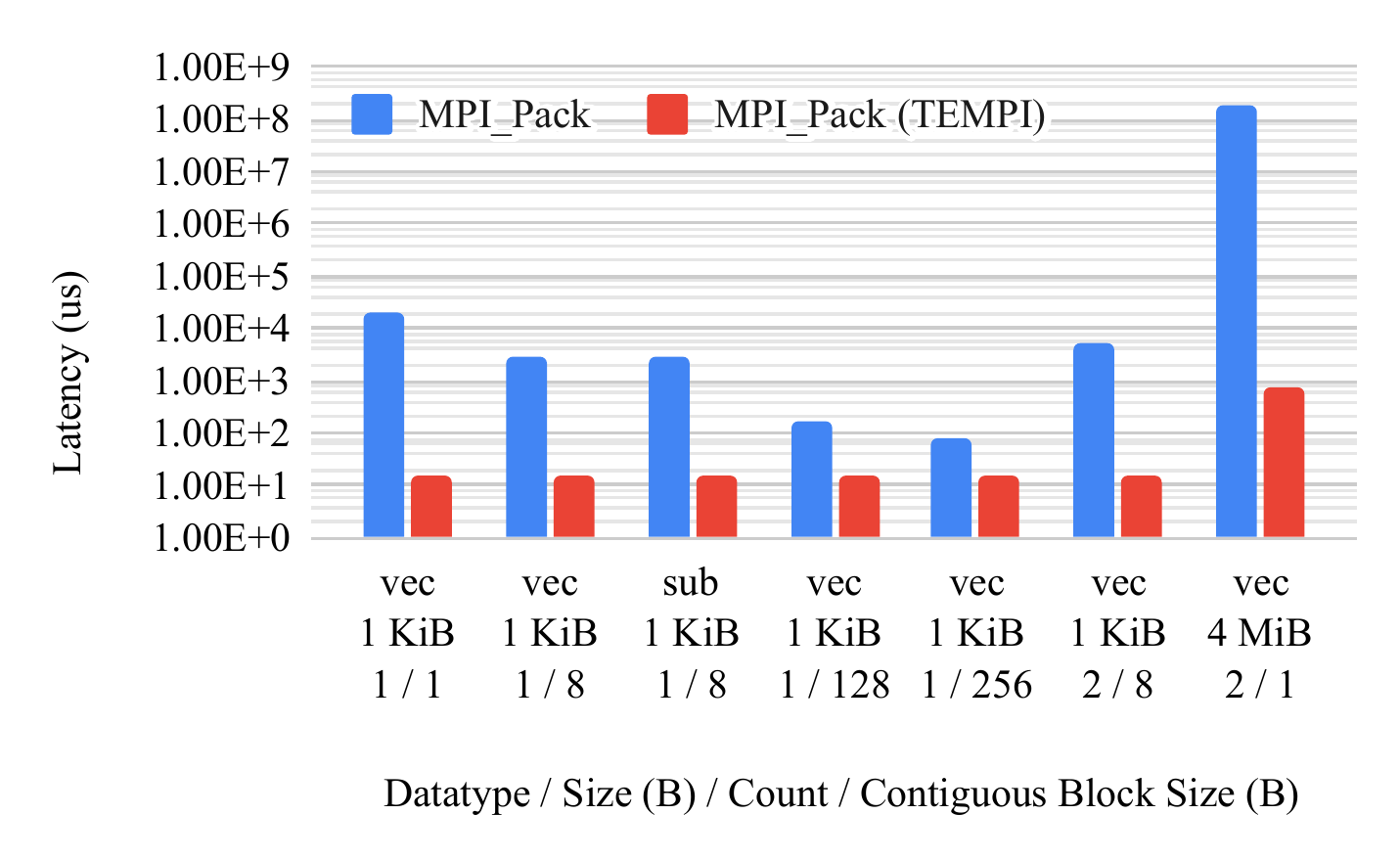}
\caption{
MPI\_Pack performance for a variety of 2D objects described as a vector or subarray datatype.
``Size'' is the total object size, ``count'' is the number of objects packed, and ``contiugous block size'' is the number of contiguous bytes in each block of the object.
The pitch of each contiguous block is \SI{512}{\byte}.
TEMPI provides fast handling regardless of MPI datatype construction or datatype count.
}
\label{fig:pack}
\end{figure}

Fig.~\ref{fig:pack} shows the pack bandwidth achieved for various 2D objects, described as a vector or subarray datatype.

Spectrum MPI 10.3.1.2 provides a baseline derived datatype handling approach where each contiguous portion of the derived datatype is copied into a contiguous buffer through cudaMemcpyAsync (or similar function).
This approach becomes faster as the contiguous block is longer (amortizing overhead), and slower with more contiguous blocks comprise the datatype.
TEMPI delivers speedup of over \num{242000} on Summit for the largest datatype.
Across the experiments speedup varies from $5.7\times$ to \num{242000}$\times$. 
Generally TEMPI performs comparatively better when the contiguous regions are smaller or the total data is larger.
In the first case, more memory copies are replaced by a single kernel, and in the second case, the GPU resources are better utilized by the kernel.
Regardless of datatype count, the MPI datatype used to provide the description, the size of the non-contiguous block, or the total size of the data, TEMPI is able to transparently provide enormous speedup.

\subsection{MPI\_Send and MPI\_Recv}
\label{sec:send}

MPI allows datatypes to be directly used with MPI\_Send and other communication routines.
Section~\ref{sec:model} introduced three ways of building MPI communication primitives with fast datatype support when the system MPI does not have it.
This model identifies when the one-shot or the staged methods are preferable.
All experiments are limited to the Summit platform, the only evaluation platform with multiple GPUs and multiple nodes.

The performance model is analyzed with measured data of various primitives.
\begin{itemize}
\item $T_{cpu-cpu}$: MPI\_Send/MPI\_Recv on CPU buffer
\item $T_{gpu-gpu}$: MPI\_Send/MPI\_Recv on GPU buffer
\item $T_{d2h}$: cudaMemcpyAsync from device (GPU) to host (CPU) and cudaStreamSynchronize
\item $T_{h2d}$: cudaMemcpyAsync from host to device and cudaStreamSynchronize
\end{itemize}
The MPI operations are measured through a ping-pong between two ranks, and the reported time is half of the total ping-pong time.
The two ranks are on separate nodes.
The CUDA operations are recorded using wall-time around the first and last calls, which reflect when control leaves and returns to the application.

\begin{figure}[htbp]
     \centering
     \begin{subfigure}[t]{\linewidth}
         \centering
         \includegraphics[width=\textwidth]{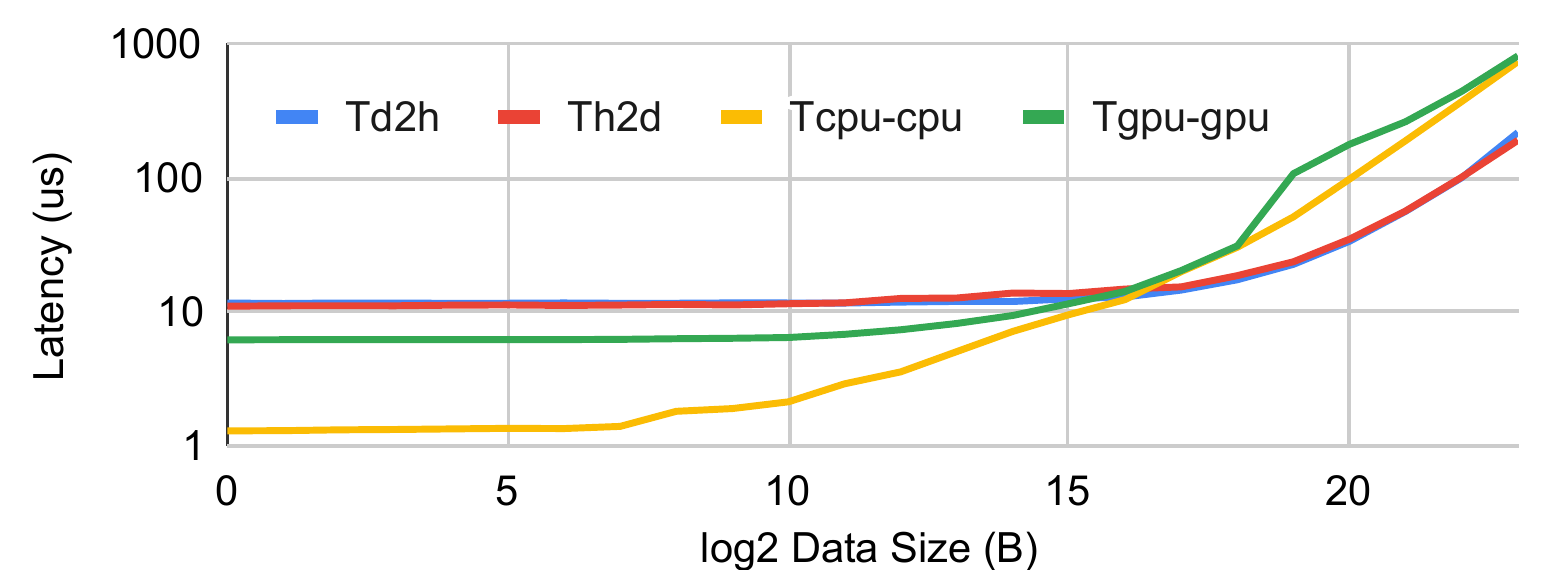}
         \caption{
Measurements of $T_{d2h}$, $T_{h2d}$, $T_{cpu-cpu}$, and $T_{gpu-gpu}$ on Summit.\newline
         }
         \label{fig:summit-raw}
     \end{subfigure}
     \hfill
     \begin{subfigure}[t]{\linewidth}
         \centering
         \includegraphics[width=\textwidth]{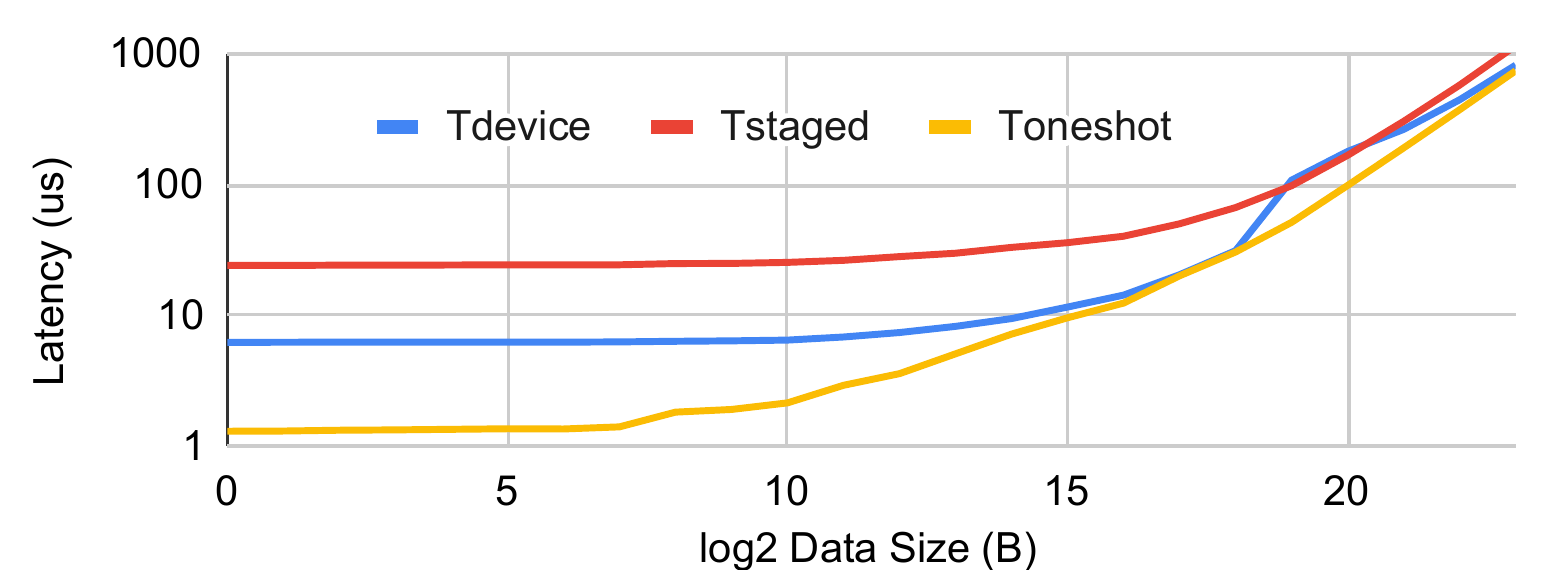}
         \caption{
Partial values of $T_{device}$, $T_{oneshot}$, and $T_{staged}$, (excluding pack time), using the values from (a).
         }
         \label{fig:summit-model-no-pack}
     \end{subfigure}
     \caption{
Raw measurements and partial performance models (omit pack/unpack) for various data transfer methods on OLCF Summit.
}
     \label{fig:summit-model}
\end{figure}

Fig.~\ref{fig:summit-raw} shows the results of the four operations for various data sizes.
CUDA-aware MPI transfers show a latency floor of approximately \SI{6}{\micro\second}, compared to \SI{1.3}{\micro\second} transfers from pinned system memory.

Fig.~\ref{fig:summit-model-no-pack} shows the measurements in Eqs.~\ref{eq:device},~\ref{eq:oneshot},~\ref{eq:staged} while holding $T_{gpu-pack/unpack}$ and $T_{cpu-pack/unpack}$ to zero (i.e., $T_{oneshot} = T_{cpu-cpu}$ and $T_{device} = T_{gpu-gpu}$).
There is no region where $T_{staged}$ is faster than $T_{device}$ and it will be disregarded for from further discussion.
Whether $T_{device}$ or $T_{oneshot}$ is faster will depend on the relative pack/unpack performance of the two methods.
As $T_{device}$ has pack/unpack occur in the faster device memory it may be faster that $T_{oneshot}$ for various transfer sizes.

\begin{figure*}[htbp]
     \centering
     \begin{subfigure}[t]{0.28\textwidth}
         \centering
         \includegraphics[width=\textwidth]{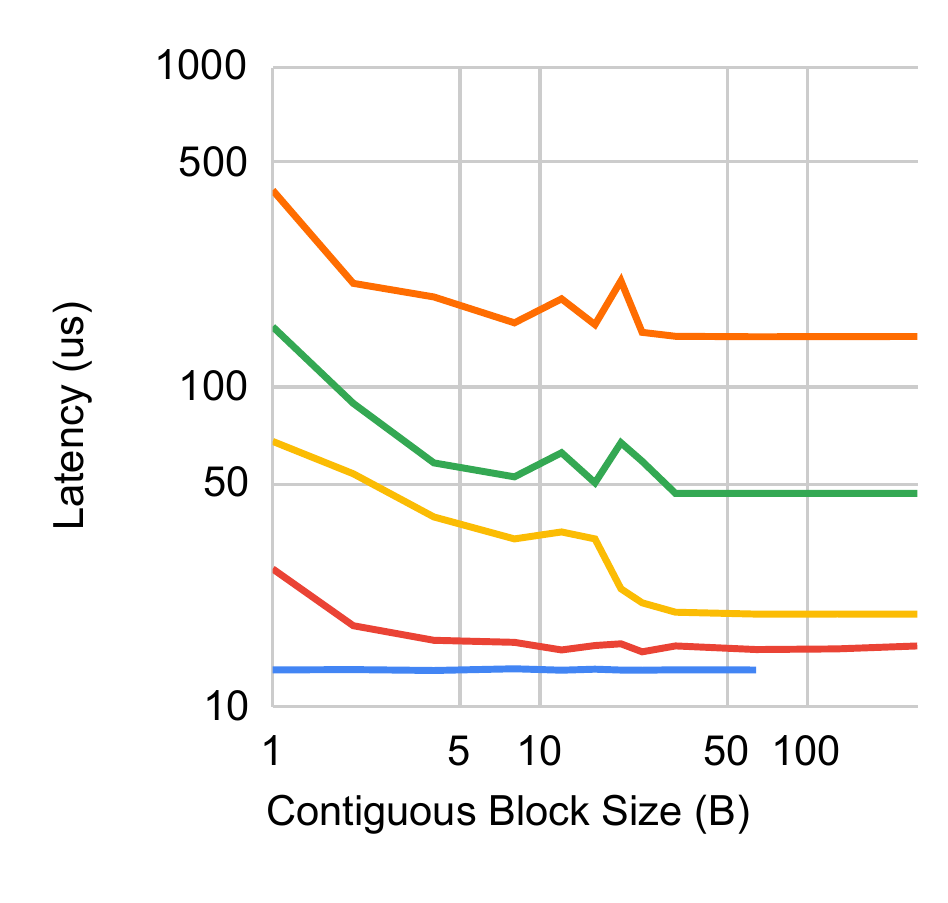}
         \caption{
One-shot pack.
         }
         \label{fig:summit-oneshot-pack}
     \end{subfigure}
     \hfill
     \begin{subfigure}[t]{0.206\textwidth}
         \centering
         \includegraphics[width=\textwidth]{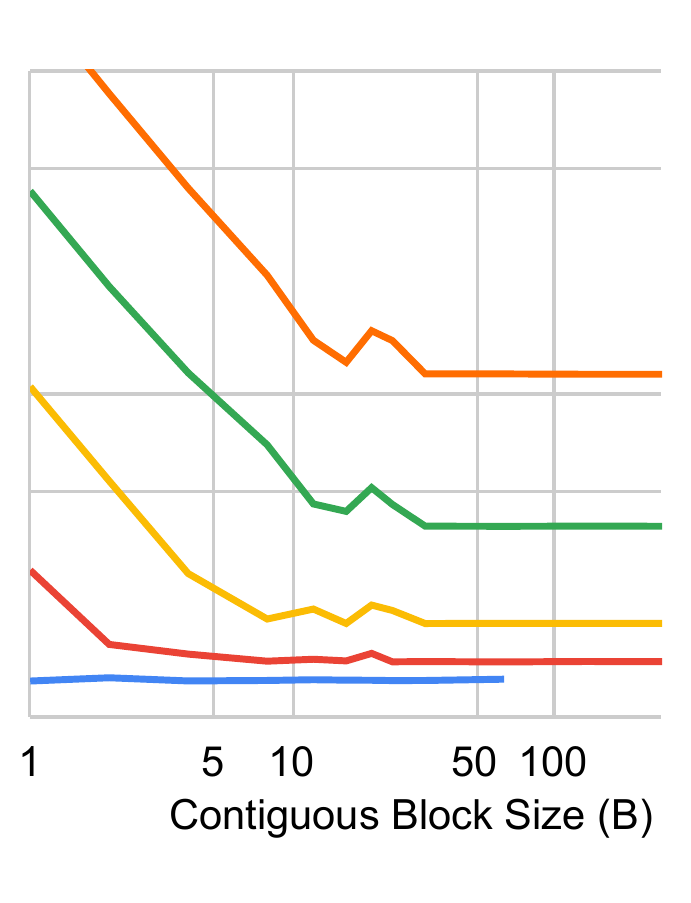}
         \caption{
One-shot unpack.
         }
         \label{fig:summit-oneshot-unpack}
     \end{subfigure}
     \hfill
     \begin{subfigure}[t]{0.206\textwidth}
         \centering
         \includegraphics[width=\textwidth]{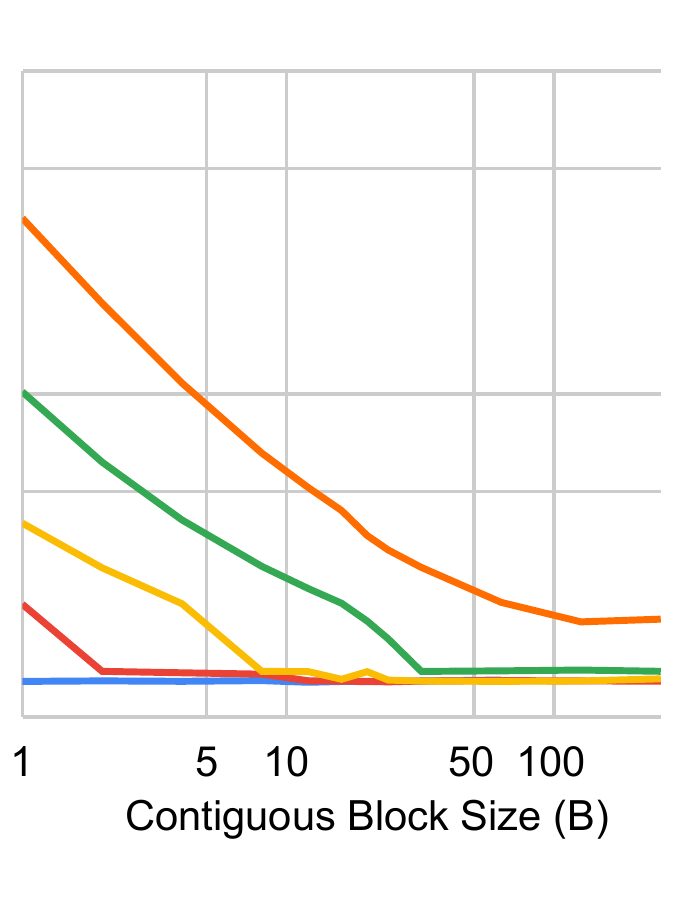}
         \caption{
Device pack.
         }
         \label{fig:summit-device-pack}
     \end{subfigure}
     \hfill
     \begin{subfigure}[t]{0.293\textwidth}
         \centering
         \includegraphics[width=\textwidth]{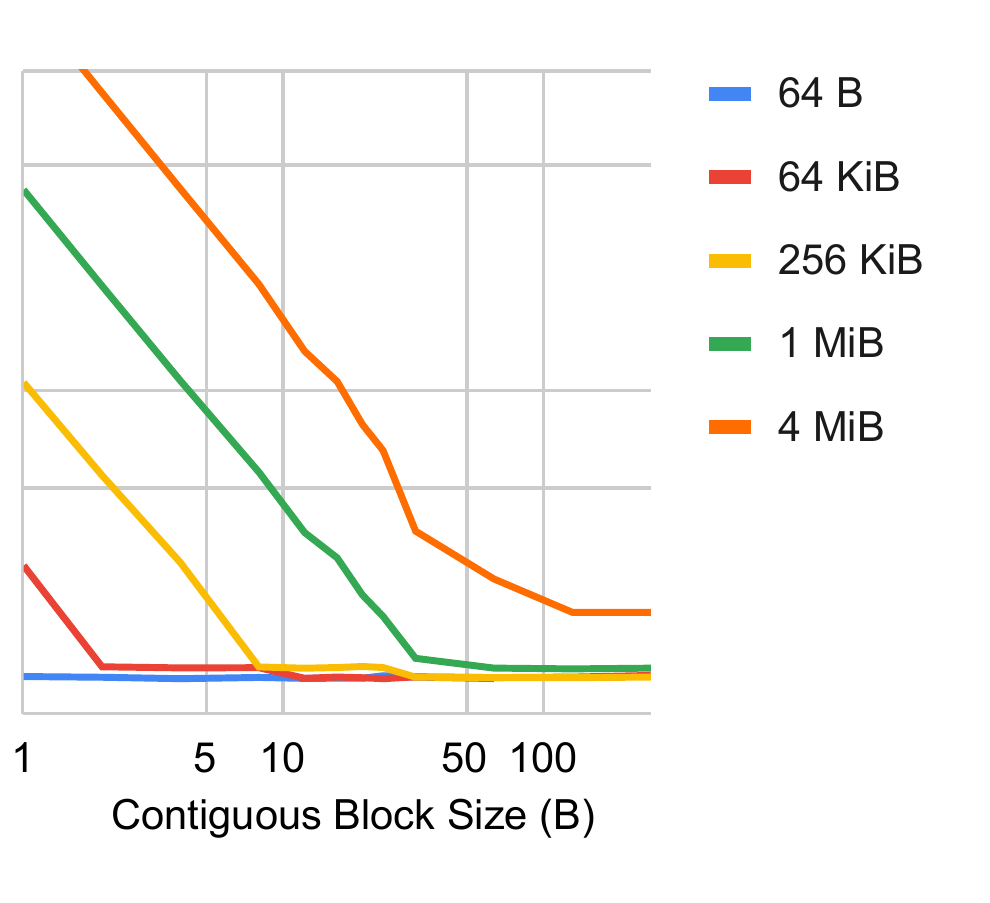}
         \caption{
Device unpack.
         }
         \label{fig:summit-device-unpack}
     \end{subfigure}
          \caption{
Pack/unpack latency using the ``one-shot'' and ``device'' strategies for \SI{64}{\byte} - \SI{4}{\mebi\byte} objects.
For smaller contiguous regions, performance is reduced due to low memory or interconnect efficiency for non-coalesced accesses.
For larger objects, performance increases as GPU resources are better utilized.
}
     \label{fig:summit-pack}
\end{figure*}

To complete the model, Fig.~\ref{fig:summit-pack} shows the measured latency of pack and unpack operations for ``one-shot'' ($T_{cpu-pack}$, $T_{cpu-unpack}$) and ``device'' ($T_{gpu-pack}$, $T_{gpu-unpack}$).
The recorded time includes all of the operations described in Section~\ref{sec:kernel-selection}, i.e. selecting appropriate grid dimensions, executing the kernel, and synchronizing after execution.


Pack/unpack latency depends on both the object size and the size of the contiguous blocks in the object.
Larger objects are faster as GPU resources are more fully utilized.
Larger contiguous blocks tend to be faster as accesses become more coalesced and make better use of memory and interconnect transactions.
One-shot performance is maximized at \SI{32}{\byte} contiguous blocks and in-device performance at \SI{128}{\byte}.
The unpack operation is slower than the pack due to non-contiguous writes (instead of non-contiguous reads in the pack operation).

Therefore, whether $T_{oneshot}$ or $T_{device}$ is faster depends on both the object size and the length of the contiguous blocks that make up that object.
Qualitatively, the one-shot method is faster when objects are smaller, as the packing kernels are limited by launch and synchronization overhead and the CPU-CPU transfers are faster than GPU-GPU.
It is also faster when objects are more contiguous, where the zero-copy accesses over the interconnect make good use of the interconnect bandwidth.

When a communication primitive is called, TEMPI uses the object size and parameters to query the performance model.
TEMPI provides a binary that records system performance parameters to the file system.
This binary should be run once before TEMPI is used in an application.
Performance measurements are sparse by necessity.
$T_{cpu-cpu}$ and $T_{gpu-gpu}$ are estimated through 1D interpolation of the object size, while $T_{cpu-pack}$, $T_{cpu-unpack}$, $T_{gpu-pack}$, and $T_{gpu-unpack}$ from a 2D interpolation from the stride and block length of the datatype.
These modeling functions are ``pure'', and their results are cached so that future invocations using the same parameters to not require a redundant expensive interpolation.

\begin{figure*}
     \centering
     \begin{subfigure}[b]{\textwidth}
         \centering
         \includegraphics[width=\textwidth]{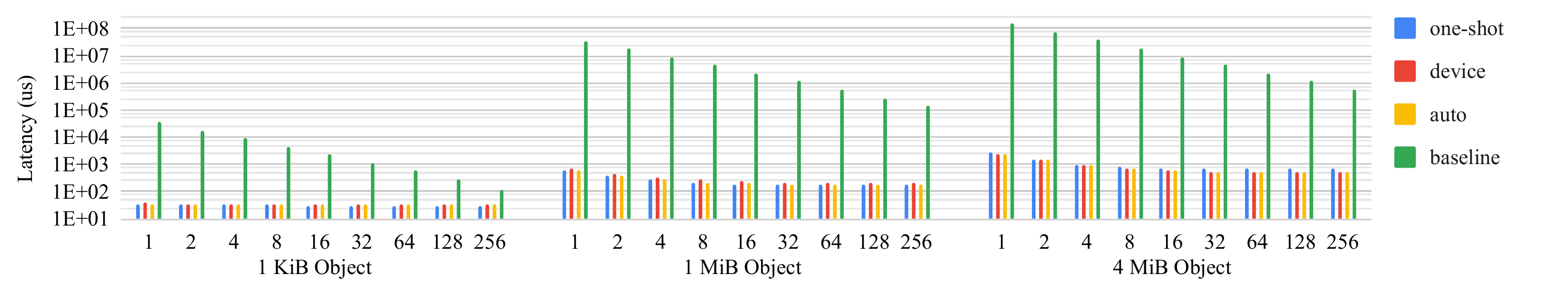}
         \caption{
MPI\_Send / MPI\_Recv latency for the one-shot, device, model-based automatic selection, and Summit MPI baseline.
The vast majority of the performance improvement comes from the datatype handling, before the one-shot or device method is selected.
         }
         \label{fig:summit-2d-send-latency}
     \end{subfigure}
     \hfill
     \begin{subfigure}[b]{\textwidth}
         \centering
         \includegraphics[width=\textwidth]{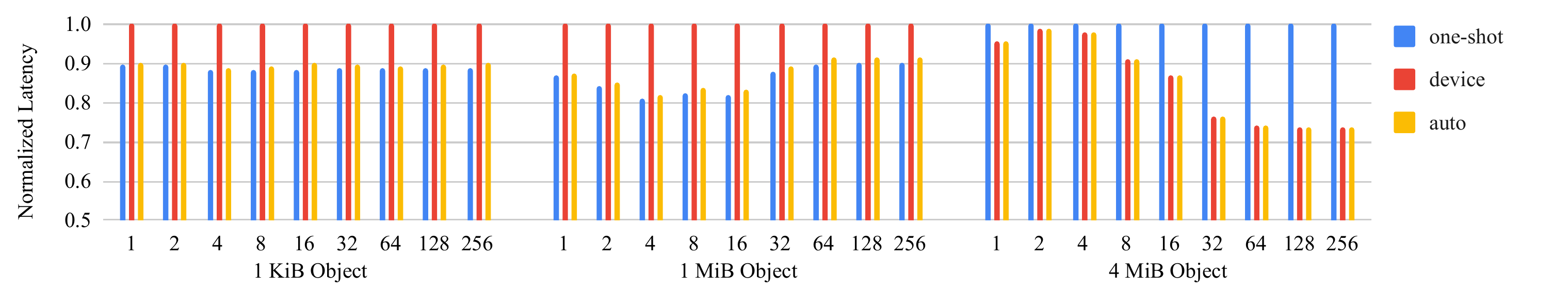}
         \caption{
Normalized latency of the one-shot, device, and model-based selection based on the measured system parameters.
The model-based automatic selection reliably chooses the faster method with minimal overhead.
         }
         \label{fig:summit-2d-send-normalized}
     \end{subfigure}
          \caption{
Time for an MPI\_Send/MPI\_Recv pair for 1KiB, 1MiB, and 4MiB 2D objects with contiguous blocks of various sizes.
``Baseline'' is the Summit platform without TEMPI.
Each group of bars is labeled with the contiguous block size.
}
     \label{fig:summit-2d-send}
\end{figure*}

Fig.~\ref{fig:summit-2d-send} shows the application-visible performance of MPI\_Send / MPI{\-}\_{\-}Recv compared to the baseline Spectrum MPI 10.3.1.2.
Fig.~\ref{fig:summit-2d-send-latency} shows that the vast majority of the speedup comes from the datatype handling (``baseline'' vs. ``one-shot''/``device'').
Since $T_{oneshot}$ or $T_{device}$ may be faster depending on the arguments passed to MPI\_Send, TEMPI uses the performance model and system measurements to estimate which method will have lower latency.
Fig.~\ref{fig:summit-2d-send-normalized} shows that the automatic model-based selection is accurate enough to reliably choose the faster of the one-shot or device methods.
In the \SI{1}{\kibi\byte} object some small model slowdown is observed as TEMPI must dynamically query the performance model to make its method selection.

This slowdown is present at all sizes, but not as visible at the larger object sizes.
Over these tests, model selection added \SI{277}{\nano\second} of latency.
The latency floor is around \SI{30}{\micro\second}, of which \SI{26}{\micro\second} can be directly attributed to the pack/unpack kernels on the sending and receiving side.
The rest of time is consumed by looking up the cached datatype handler and checking to see if the user-provided buffers are GPU-resident.
Speedup between the baseline and TEMPI's automatic selection is up to $59000\times$ for large objects with small blocks.

\subsection{Case Study: 3D Stencil}
\label{sec:stencil}

The enormous datatype handling performance has a commensurate impact on application performance.
Here we consider a 3D stencil code, where the total stencil region is a cube of $256^3 \times P$ gridpoints and $P$ is the number of ranks.
Each gridpoint has eight eoght-byte value, and the stencil radius is $3$.
This implementation is chosen to replicate the communication requirements of the Astaroth stellar simulation code \cite{pekkila2017methods}.
Each halo region is defined in a separate MPI derived datatype and packed into the single buffer using MPI\_Pack.
Halo exchange is implemented as an MPI\_Alltoallv on that single buffer.
Then, the receive buffer is unpacked.
The stencil kernel is a standard 26 point, yielding 26 neighbors for each rank with periodic domain boundaries.
This means each rank engages in 26 MPI\_Pack and 26 MPI\_Unpack operations on a variety of different 3D strided datatypes.
Reported times are the maximum time consumed for each phase across all ranks.

\begin{figure*}[ht]
\centering

     \begin{subfigure}[t]{0.49\textwidth}
         \centering
         \includegraphics[width=1.0\textwidth]{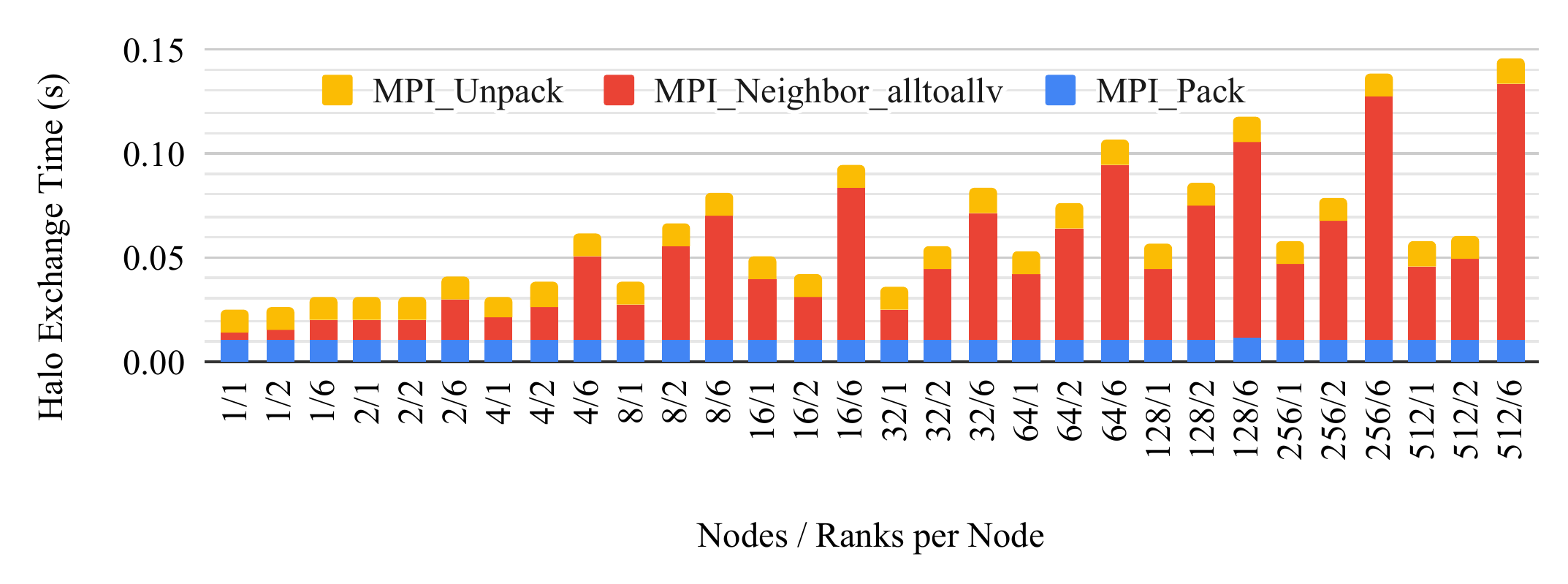}
         \caption{
Performance of phases of halo exchange.
Alltoallv time is higher for more ranks and more nodes.
Pack/Unpack time is constant, as the amount of data moved per rank is unchanged.
         }
         \label{fig:halo_latency}
     \end{subfigure}
     \hfill
     \begin{subfigure}[t]{0.49\textwidth}
         \centering
         \includegraphics[width=1.0\textwidth]{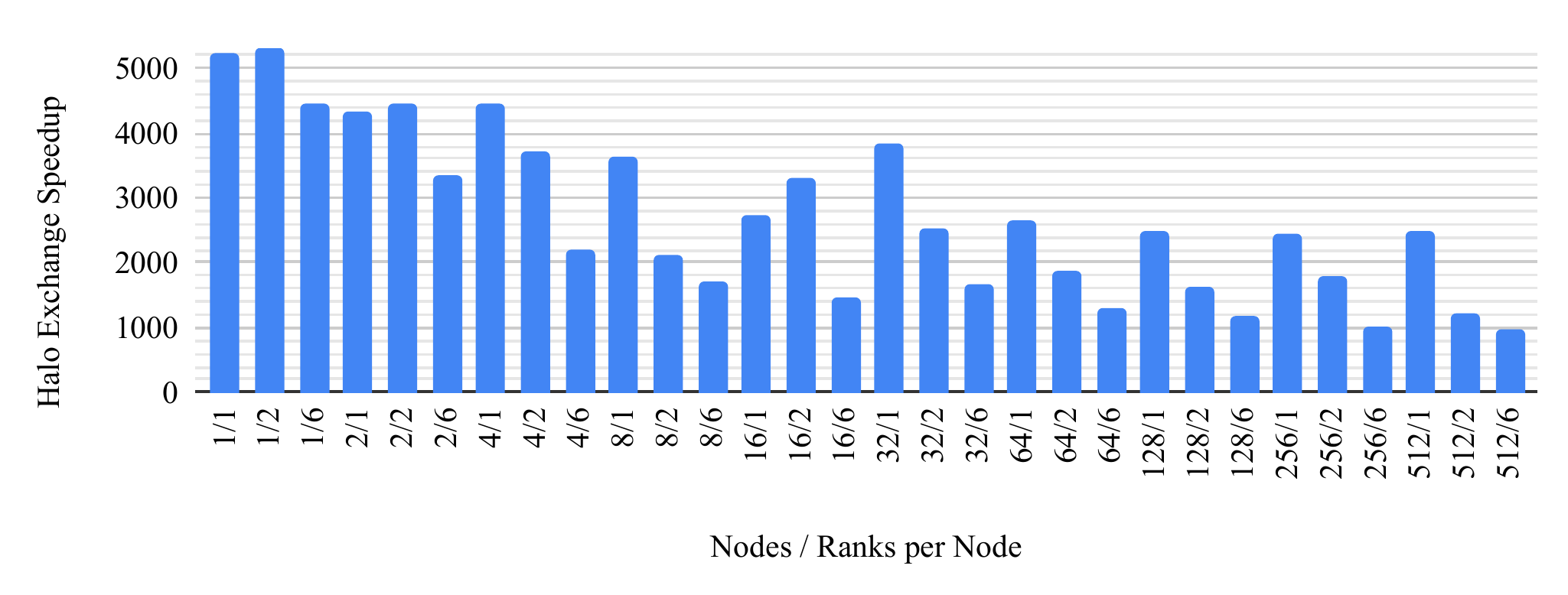}
         \caption{
Speedup of entire halo exchange.
Speedup is lower for larger number of ranks as datatype handling is a smaller portion of the total time.
         }
         \label{fig:halo_sp}
     \end{subfigure}

\caption{
Performance of 3D stencil halo exchange operation using TEMPI on Summit with SpectrumMPI 10.3.1.2.
}
\label{fig:stencil-results}
\end{figure*}

Fig.~\ref{fig:stencil-results} shows TEMPI's halo exchange latency on Summit, broken down into pack, MPI\_Alltoallv, and unpack, as well as the overall halo exchange speedup.
The different pack and unpack latencies reflects their different kernel performance (Section~\ref{sec:pack}).
The speedup is smallest for larger number of ranks, as the communication takes up a relatively larger amount of the total iteration time compared to the pack and unpack operations.
Speedup at 192 ranks is $1050\times$.

\section{Related Work}
\label{sec:related}

\begin{table*}[ht]
\centering
\caption{
Selected microbenchmark results from related work.
Provided for reference, as hardware changes hinder direct performance comparison.
}
\label{tab:related}
\resizebox{\textwidth}{!}{%
\begin{tabular}{c|c|c|c|ccc}
\multirow{2}{*}{\textbf{Work}} & \multirow{2}{*}{\textbf{Platform}} & \multicolumn{2}{c|}{\textbf{Non-contiguous Operation}}                 & \multicolumn{3}{c}{\textbf{Nominal Bandwidth Slowdown}}                                                    \\ \cline{3-7} 
                               &                                    & \textbf{Pack}                & \textbf{Dis. Mem. Ping-Pong}                      & \multicolumn{1}{c|}{\textbf{Network}} & \multicolumn{1}{c|}{\textbf{GPU Mem.}} & \textbf{GPU Interconnect} \\ \hline
\cite{wang2011optimized}              & C2050, QDR IB                      & 25us (1KiB ), 10ms (4MiB)    & 20ms (4MiB)                      & \multicolumn{1}{c|}{3.125}            & \multicolumn{1}{c|}{6.25}              & 6.25                      \\
\cite{shi2014hand}                    & C2050, QDR IB                      & 120us (1KiB)                 &    \textit{(none provided)}         & \multicolumn{1}{c|}{3.125}            & \multicolumn{1}{c|}{6.25}              & 6.25                      \\
\cite{jenkins2014processing}         & C2050, QDR IB                      & 10us (1 KiB)                 & 70us (1KiB), 700us (256KiB)             & \multicolumn{1}{c|}{3.125}            & \multicolumn{1}{c|}{6.25}              & 6.25                      \\
\cite{wei2016gpuaware}                & K40, FDR IB                        & 75us (512KiB), 150us (4 MiB) & 7ms (4 MiB)                             & \multicolumn{1}{c|}{1.8}              & \multicolumn{1}{c|}{3.125}             & 3.175                     \\
This                           & V100, EDR IB                       & 13us (64 KiB), 21us (4 MiB)  & 60us (1KiB), 354us (1MiB), 888us (4MiB) & \multicolumn{1}{c|}{1}                & \multicolumn{1}{c|}{1}                 & 1                        
\end{tabular}
}
\end{table*}

TEMPI distinguishes itself from prior work in three ways.
First, TEMPI can be used today without waiting for MPI implementers or HPC system administrators.
Second, while prior work uses GPU kernels to accelerate datatype operations, TEMPI is the first work that shows transformations on structured datatypes for canonicalization (as opposed to handling specific cases, or reducing everything to a list of offsets and lengths).
This provides wide datatype coverage, tiny GPU memory consumption for metadata, and fast generic kernels.
Third, prior work examines how to integrate data type handling into MPI more deeply, including pipelining and zero-copy transfer between GPUs.
As TEMPI uses a library-interposer interface on top of MPI, it is not able to make those low-level changes.
Despite that, enormous performance improvement was achieved.

The MPITypes library~\cite{ross2009mpitypes} is one of the first attempts to generalize datatype handling outside of MPI.
It provides several functions for flattening and copying datatypes, and a framework for extending those operations.
TEMPI tries to maintain the structured information of types so the MPITypes operations are not directly applicable.

Wang et al.~\cite{wang2011optimized} describes an early approach in MVAPICH2.
Several options are considered, ultimately selecting a pipelined version of the ``staged'' method that uses cudaMemcpy2D instead of a kernel.


Jenkins et al. \cite{jenkins2014processing} provide fast handling of arbitrary MPI datatypes on the GPU.
Nested types are represented by a tree structure that must be traversed by each GPU thread using division, modulo, and binary search operations.
They restrict inter-node communication to the one-shot method, which this work finds is not always preferable.


Shi et al.~\cite{shi2014hand} also explicitly breaks the problem into transformation and kernel selection phase.
Hand defines specific kernels for handling vector, hvector, subarray, and indexed block types.
For  other datatypes, it transforms a variety of datatypes into a blocklist, for which it has a specific kernel implementation.
Instead, TEMPI recognizes that nested, strided types reduce to (essentially) a subarray, and explicitly designs a transformation and optimal packing kernel to cover all of those scenarios.

Wei et al.~\cite{wei2016gpuaware} describe a fork of OpenMPI that integrates derived datatype handling both on the GPU itself as well as communication between nodes.
The datatype is ultimately represented as a list of blocks, and blocks are partitioned among separate kernels with pipelined communication.
It also identifies that full GPU resources for handling non-contiguous data are not needed to saturate the communication links.
This fork has remained unmerged and not publicly available.

Chu et al.~\cite{chu2019high} recognize that one of the challenges of all prior work is the latency of kernel launches.
Like prior work, it also represents the datatypes as a list of displacements and lengths.
Similar to this work, it defines extraction, conversion, and caching steps, and uses one-shot packing and unpacking.
Unlike TEMPI, they do not recognize when the one-shot packing to the host is slower to inefficiency of packing and unpacking over the interconnect.

Chu et al.~\cite{chu2020dynamic} identify that a major cost of transfer is the launch of the packing kernels.
They develop an engine that is able to merge various packing requests into a single kernel launch.
TEMPI addresses the packing kernel launch cost by issuing a single kernel for multiple copies of the same MPI datatype, but cannot fuse further than that.

Hashmi et al.~\cite{hashmi2020falcon} describe a zero-copy-based data movement system for MPI datatypes.
They include kernels where a warp is responsible for a contiguous block in a block list.
They also describe integration with the underlying communication library, which TEMPI is unable to address due to its interposer model.

Yaksa \cite{yaksa2021} is a high-performance non-contiguous datatype engine being developed for MPICH.
Like this work, it features an internal representation of non-contiguous data derived from MPI datatypes.
Custom GPU kernels are synthesized for each Yaksa datatype that corresponds to the familiar MPI datatypes.
When datatypes are nested, the result is traversed and pack operations are issued separately for each subtree for which there is a synthesized kernel.

Not all prior works include microbenchmarks comparable to the ones presented herein.
Table ~\ref{tab:related} summarizes those that do.
Substantial hardware changes make a direct comparison difficult, so normalized bandwidth numbers for various subsystems are provided for reference.
TEMPI is competitive for both latency- and bandwidth-bound operations (small and large amounts of data, respectively). 

\section{Future Work}
\label{sec:future}

Several other MPI implementations, including MVAPICH (and especially MVAPICH-GDR), Open MPI, and MPICH, have varying degrees of support for fast GPU datatype handling.
As these implementations are not available on Summit, a direct comparison was not possible.
However, TEMPI's approach is complementary to what is adopted by those implementations.
TEMPI could be used to prototype improvements or extensions to the currently-implemented datatype handling, just as it was for Spectrum MPI.
Furthermore, prior work that has not been integrated with an existing MPI implementation could use TEMPI to evaluate and deployment to a large variety of existing systems and codes.
TEMPI could also be extended to use an approach from prior works to handle indexed datatypes (including MPI\_Type\_struct) on Summit, or to evaluate the use of the GPU DMA engine for non-contiguous data (e.g. cudaMemcpy2D).

The performance modeling and runtime implementation choice is currently limited to MPI\_Send and MPI\_Recv pairs.
Furthermore, the interposer library model hinders deeper modification of MPI primitives, such as pipelining packing and interprocess communication.
Even simple asynchronous operations (MPI\_Isend) introduce additional modeling and measurement challenges, as many communications can be active simultaneously.
Similarly, this work does not address MPI collectives.
Bienz et al.~\cite{bienz2018improving} have introduced some performance modeling techniques that could be extended to evaluate the impact of improved datatype handling in that context.

\section{Conclusion}
\label{sec:conclusion}
Despite years of deployment of CUDA-aware MPI systems alongside research contributions for GPU datatype handling, fast MPI derived datatype handling is not available on all platforms.
This work used a library-interposer approach for deploying fast datatype handling on OLCF Summit without requiring modification to the system or applications.
It presented a novel approach for transforming datatypes describing common objects into a compact canonical form, backed by generic processing kernels.
Empirical system measurements are used at run-time to optimize the packing method.
MPI\_Pack performance on datatypes was sped up by up to \num{242000}$\times$, MPI\_Send up to \num{59000}$\times$ and a 3D stencil halo exchange up to \num{917}$\times$ at $3072$ processes.
\ifdefined\BLIND
\else
TEMPI is available at \url{https://github.com/cwpearson/tempi}.
\fi

\begin{acks}
This work is supported by IBM-ILLINOIS Center for Cognitive Computing Systems Research (C3SR) - a research collaboration as part of the IBM AI Horizon Network.
This research used resources of the Oak Ridge Leadership Computing Facility at the Oak Ridge National Laboratory, which is supported by the \grantsponsor{DE-AC05-00OR22725}{Office of Science of the U.S. Department of Energy}{} Contract No. \grantnum{DE-AC05-00OR22725}{DE-AC05-00OR22725}.
This work utilizes resources supported by the \grantsponsor{1725729}{National Science Foundation}{}'s Major Research Instrumentation program, grant \#\grantnum{1725729}{1725729}, as well as the University of Illinois at Urbana-Champaign.
The authors would also to thank Dawei Mu, Omer Anjum, and Mert Hidayetoglu.
\end{acks}

\bibliographystyle{ACM-Reference-Format}
\bibliography{main}


\end{document}